\documentclass[11pt,letterpaper]{article}
\usepackage{./files/jheppub}
\usepackage{amsmath}
\usepackage{amssymb}
\usepackage{color}
\usepackage{graphicx}
\usepackage[utf8x]{inputenc}
\usepackage{default}
\usepackage{mathtools}
\usepackage{array}
\usepackage{mathrsfs}
\usepackage{multirow}

\renewcommand{\arraystretch}{1.5}

\graphicspath{{./figures/}}

\newcommand{\beq}{\begin{equation}}
\newcommand{\eeq}{\end{equation}}
\newcommand{\bea}{\begin{eqnarray}}
\newcommand{\eea}{\end{eqnarray}}
\newcommand{\C}{\mathbb{C}}

\newcommand{\R}{\mathbb{R}}

\newcommand{\trop}{{\text{trop}}}
\newcommand{\mon}{\mathop{\rm mon}}
\newcommand{\Cs}{\mathcal{V}}
\newcommand{\W}{\widetilde{\mathcal{W}}}
\renewcommand{\d}{\partial}
\newcommand{\dilog}{\mathop{\rm Li}_2\nolimits}
\renewcommand{\Re}{\mathop{\rm Re}\nolimits}
\renewcommand{\Im}{\mathop{\rm Im}\nolimits}
\newcommand{\vs}{\vspace{0.5ex}}

\def\tilde{\widetilde}
\def\hat{\widehat}
\def\bar{\overline}

\def\CB{{\mathcal B}}

\def\CF{{\mathcal F}}

\def\CI{{\mathcal I}}

\def\CM{{\mathcal M}}
\def\CN{{\mathcal N}}
\def\CO{{\mathcal O}}

\def\CT{{\mathcal T}}

\def\CV{{\mathcal V}}
\def\CW{{\mathcal W}}

\title{Walls, Lines, and Spectral Dualities in 3d Gauge Theories}
\preprint{CALT 68-2905}
\author{Abhijit Gadde, Sergei Gukov, and Pavel Putrov}
\affiliation{California Institute of Technology \\ Pasadena, CA 91125, USA}
\emailAdd{abhijit@caltech.edu}
\emailAdd{gukov@theory.caltech.edu}
\emailAdd{putrov@theory.caltech.edu}
\bigskip

\abstract{In this paper we analyze various half-BPS defects in a general three dimensional $\CN=2$ supersymmetric gauge theory $\CT$. They correspond to closed paths in SUSY parameter space and their tension is computed by evaluating period integrals along these paths. In addition to such defects, we also study wall defects that interpolate between $\CT$ and its $SL(2,{\mathbb Z})$ transform by coupling the 3d theory to a 4d theory with S-duality wall. We  propose a novel spectral duality between 3d gauge theories and integrable systems. This duality complements a similar duality discovered by Nekrasov and Shatashvili. As another application, for 3d $\CN=2$ theories associated with knots and 3-manifolds we compute periods of (super) $A$-polynomial curves and relate the results with the spectrum of domain walls and line operators.}

\makeatother

\begin{document}
\maketitle

\section{Defects and periods in 3d} \label{sect_defects}

The soliton structure contains a lot of useful information about $\CN=2$ supersymmetric theories in two dimensions \cite{FMVW,FLMW,LercheW} and, in fact, even offers a way to classify such theories and their massive deformations~\cite{CVclassification}. The central role in this framework is played by a superpotential function that we denote $\tilde \CW$ and whose critical points correspond to the massive vacua of the deformed theory.

In the present paper we tackle a similar problem for $\CN=2$ theories in three dimensions, which upon reduction on a circle also lead to 2d $\CN=(2,2)$ theories. Again, the central role will be played by a (twisted) superpotential function $\tilde \CW$. However, for theories obtained via Kaluza-Klein reduction from three dimensions, the function $\tilde \CW$ in general will have infinitely many critical points, resulting in an infinite spectrum of supersymmetric vacua and solitons that interpolate between them. This is one of the features that makes 3d theories on $S^1_t \times \R^2$ very rich compared to typical 2d $\CN=2$ theories with finitely many vacua. Another reason is that solitons --- which in 2d are the same as domain walls --- in the Kaluza-Klein reduction on a circle can originate from two-dimensional as well as one-dimensional defects in 3d.

Our goal is to probe the rich dynamics of 3d $\CN=2$ gauge theories by studying such defects, specifically line operators and domain walls. Since domain walls in three dimensions are supported on two-dimensional surface, they can also be called ``surface operators.'' We shall see the role of these defects in the calculation of various partition functions, {\it e.g.} the supersymmetric index $\CI_{S^1_t \times S^2}$ on $S^1_t \times S^2$, the ellipsoid partition function $Z_{S^3_b}$ on $S^3_b$, or the vortex partition function $Z_{\text{vortex}}$ on~$S^1_t \times_q \R^2$. Here the symbol $\times_q$ denotes the twisted product with equivariant parameter $q$.

The latter partition function plays the role of a basic building block \cite{DGGindex,Pasquetti} since many three-dimensional space-times can be constructed by gluing copies of $S^1_t \times_q D$,where $D \cong \R^2$ is a ``cigar.'' For this reason, it was called a 3d ``holomorphic block'' in the recent work \cite{BDP}. In particular, the index $\CI_{S^1_t \times S^2} (\CT)$ of a theory $\CT$ can be built from two copies of the ``half-index'' $\CI_{S^1_t \times_q D} (\CT) \cong Z_{\text{vortex}} (\CT)$ on $S^1_t \times_q D$, where $D$ can be identified with the disk covering either upper or lower hemisphere of the $S^2$.

Denoting by $S^1_{sp}$ the equator of $S^2$ in the index calculation or, equivalently, the boundary of the disk, $S^1_{sp} := \partial D$, in the half-index / holomorphic block calculation we shall consider the following configurations of line and surface operators:

\begin{itemize}
\item
line operators supported on the ``temporal circle'' $S^1_t$;
\item
line operators supported on the ``spatial circle'' $S^1_{sp}$;
\item
domain walls supported on $S^1_t \times S^1_{sp}$.
\end{itemize}

\begin{figure}
\center{\includegraphics[width=5in]{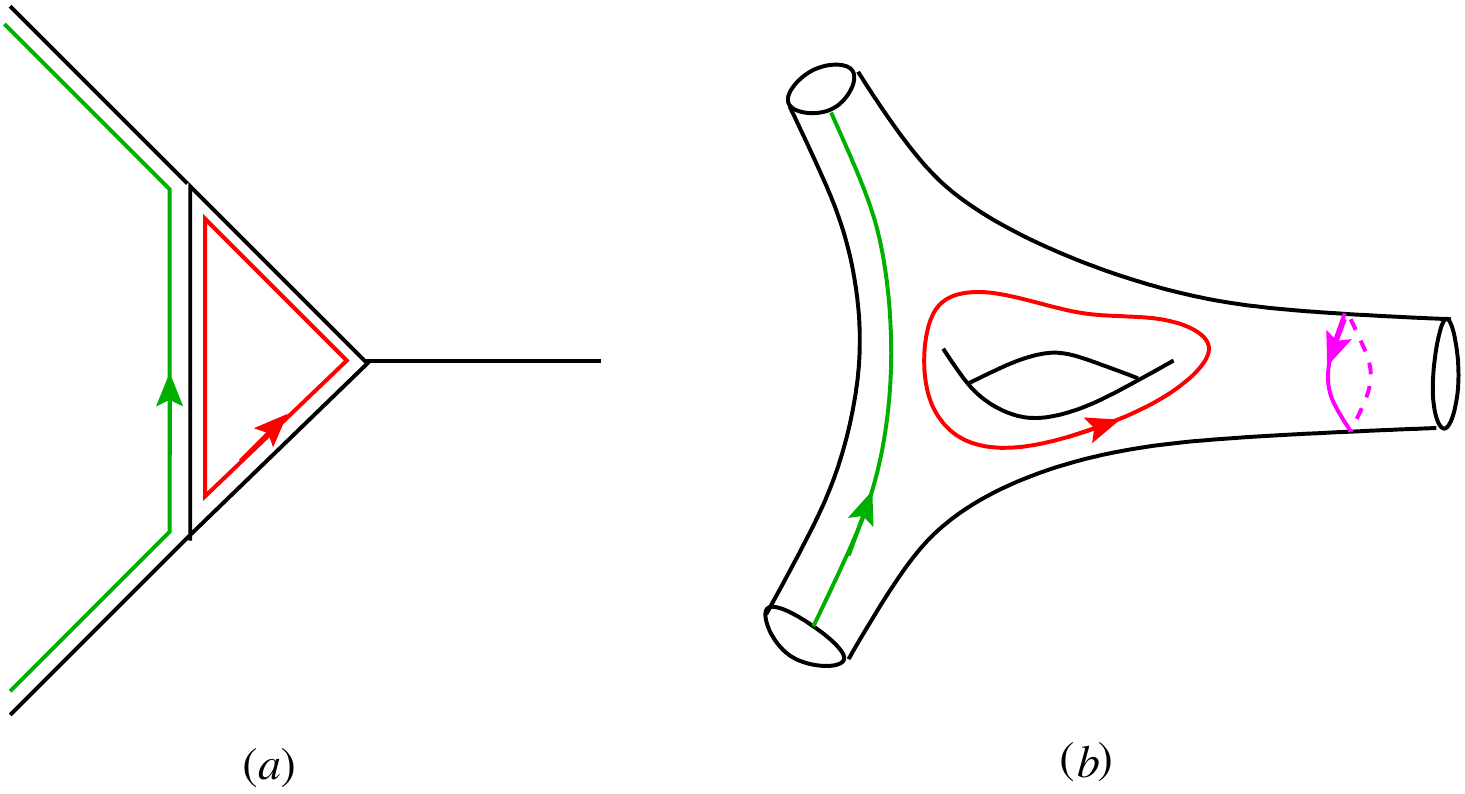}}
\caption{Different types of paths on the parameter space of 3d $\CN=2$ theory and its circle compactification.
Note, that closed cycles that go around asymptotic regions of the curve $\CV$ disappear in the 3d / tropical limit.
\label{fig:curve2}}
\end{figure}

In particular, as we explain in what follows, some of these defects have a simple and natural interpretation in terms of periods on an algebraic curve or, more generally, an algebraic variety
\beq
\CV \; \subset \; \left( \C^* \times \C^* \right)^n
\label{VinCC}
\eeq
associated with the 3d $\CN=2$ theory $\CT$ with $n$ global $U(1)$ flavor symmetries.\footnote{More generally, for a 3d theory with a non-abelian global symmetry $G$, $\CV$ is a subvariety of $(\mathbb{T} \times \mathbb{T}) / \CW_G$, where $\mathbb{T}$ is the maximal torus of the complexification, $G_{\C}$, and $\CW_G$ is the Weyl group.} The algebraic variety $\CV$ controls the asymptotic behavior of various partition functions and can also be understood as the space of supersymmetric parameters in the 3d $\CN=2$ theory $\CT$ compactified on a circle. Thus, in the limit $q \to 1$ we have
\beq
\left .
\begin{array}{l}
Z_{S^3_b} \\
Z_{\text{vortex}}=\CI_{S^1_t \times_q D} \\
\CI_{S^1_t \times S^2}
\end{array}
\right \}
\; \simeq \; \exp \left( \frac{1}{\hbar} \tilde \CW + \ldots \right)
\label{ZZZasymp}
\eeq
where $q = e^{\hbar}$ and $\tilde \CW$ can be interpreted as the twisted superpotential of the 3d $\CN=2$ theory compactified on a circle $S^1_t$. As explained in \cite{DGSdual}, the twisted superpotential $\tilde \CW$ is given by the integral over an open path on $\CV$ connecting $p\in\CV$ with some reference point $p_*$:
\beq
\tilde \CW(p) \; = \; \int_{p_*}^p \lambda
\label{Wint}
\eeq
where
\beq
\lambda \; = \; \sum_{i=1}^n \log y_i \frac{dx_i}{x_i}
\eeq
is a 1-form written in terms of the $( \C^* )^n \times ( \C^* )^n$-valued coordinates $(x,y)$ that parametrize \eqref{VinCC}. The path $\gamma$ starts at some reference point on $\CV$ (that we tacitly assume to be fixed throughout our discussion), and ends at the point $p=(x,y) \in \CV$ that represents SUSY vacuum of our interest. To be more precise, $p \in \CV$ represents the choice of parameters (twisted masses, FI terms, {\it etc.}) for which the theory $\CT$ compactified on a circle has a SUSY vacuum. Note, defined by \eqref{Wint} the twisted superpotential $\tilde \CW$ is a multi-valued function and its ambiguity is given precisely by the closed periods on $\CV$. In what follows we explain that the three types of above mentioned defects correspond to different types of monodromies
\beq
\tilde \CW \; \to \; \tilde \CW + \Delta \tilde \CW
\eeq
and some correspond to periods, as shown in Figure~\ref{fig:curve2}.

In the opposite direction, the expression \eqref{Wint} implies that the space of SUSY parameters $\CV$ is a middle-dimensional hypersurface in $( \C^* )^n \times ( \C^* )^n$ defined by $n$ equations:
\beq
y_i = \exp \left( x_i \frac{\partial \tilde \CW}{\partial x_i} \right)
\qquad i = 1, \ldots, n \,.
\label{Vxiyi}
\eeq
In fact, from this description it is clear that $\CV$ is a complex Lagrangian submanifold with respect to the holomorphic symplectic form
$\Omega = \sum_i d \log x_i \wedge d \log y_i$.
For simplicity, in what follows we mostly focus on the basic case $n=1$; the generalization to $n>1$ is completely straightforward. In that case, $\CV$ is an algebraic curve defined by a single equation that we often write as
\beq
\CV~: \quad A (x,y) \; = \; 0 \,.
\label{Acurve}
\eeq

From another perspective, this algebraic curve associated to a 3d $\CN=2$ theory $\CT$ on $S^1_t \times \mathbb{R}^2$ defines a boundary condition for a 4d $\mathcal{N}=4$ abelian gauge theory on half-space $S^1_t \times \mathbb{R}^2 \times \mathbb{R}_+$. The moduli space of vacua of this four-dimensional theory is
\begin{equation}
\mathcal{M}_{4d} \; = \; \{(x,y)\in \mathbb{C}^*\times \mathbb{C}^*\}  \times \mathbb{C}^2 \,,
\label{4d_moduli}
\end{equation}
so that $x$ and $y$ are vacuum expectations values of Maldacena-Wilson and 't Hooft loop operators supported on ${S}^1_t$, see \cite{HMS,BJSV,SW3d,Seiberg16}. The 4d theory can be understood as the theory on a M5 brane compactified on a torus \cite{Wittencomments,OVN4,Stromingeropen}. Four real dimensions of the $\mathbb{C}^2$ factor together with $(\log|x|,\log|y|)$ correspond to vacuum expectation values of 6 scalar fields. The Wilson and 't Hooft operators can be associated to two cycles of the torus \cite{Henningson,DMO,AGGTV,DGOT}, so that the electric-magnetic duality of $\CN=4$ gauge theory acts in a natural way by $SL(2,\mathbb{Z})$ matrices on the vector $(\log x,\log y)$. The boundary condition defined by a 3d theory $\CT$ reduces the moduli space \eqref{4d_moduli} to a Lagrangian submanifold $\mathcal{L}_{\Cs} \subset \mathcal{M}_{4d}$,
\begin{equation}
\begin{array}{c}
 \mathcal{L}_\Cs=\Cs\times\mathbb{C}\\
 \Cs=\{(x,y)\in \mathbb{C}^*\times \mathbb{C}^*\;|\;A(x,y)=0\}
 \label{4d_boundary_moduli}
\end{array}
\end{equation}
where $A(x,y)$ is the polynomial that we already encountered in the above discussion of SUSY vacua and partition functions of the 3d theory~$\CT$.

Since the 3d boundary theory $\CT$ is compactified on a circle ${S}^1_t$ it can be considered as a two-dimensional theory with $\CN=(2,2)$ supersymmetry. Let us consider the effective twisted superpotential $\CW (x)$ already minimized with respect to all dynamical twisted chiral multiplets. Then, the $U(1)$ gauge field of the 4d theory induces a background field for a global symmetry $U(1)_x$ in the 3d boundary theory. This field can be considered as a part of a twisted chiral superfield $\Sigma_x=\bar{D}_+D_-V_x$ in the effective 2d $\CN=(2,2)$ theory, so that $\langle \Sigma_x \rangle=\log x$ and $V_x$ is the vector superfield for $U(1)_x$ symmetry. Thus $\CW (x)$ determines how the 4d theory is coupled to the boundary 3d theory $\CT$ by gauging its global (flavor) symmetry.

\subsection{Algebra of line operators}

In \cite{ramified} it was argued that line operators supported on a surface operator can be identified with elements of the fundamental group
\beq
\{ \text{line operators} \} \; = \; \pi_1 (\CV)
\label{linesfundgp}
\eeq
where $\CV$ is the space of supersymmetric parameters of the surface operator. In particular, this description of the algebra of line operators was used in \cite{ramified} to realize the affine Hecke algebra and its categorification, the affine braid group. Even though here our context is slightly different and we are interested in systems with half as much supersymmetry, the argument (that we shall review below) is essentially the same and also leads to the conclusion \eqref{linesfundgp}.

\begin{figure}
\centering
\includegraphics{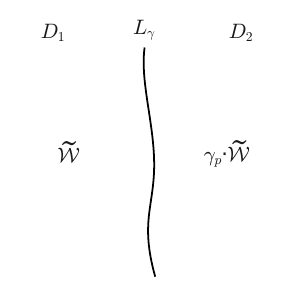}
\caption{The defect $L_\gamma$ separating two domains $D_1$ and $D_2$ of the effective 2d theory.}
\label{figure_2d-defect}
\end{figure}

One can see that the definition of the effective twisted superpotential through the integral (\ref{Wint}) is ambiguous and depends on the choice of the path connecting $p$ and $p_*$. The different choices are related by the elements of $\pi_1(\CV,p)$. This defines an action on the effective twisted superpotential:
\begin{equation}
\gamma_p\cdot\W(p) =\W(p)+\Delta_{\gamma_{p}} \W (p),\qquad \gamma_p \in \pi_1(\CV,p)
\end{equation}
where
\begin{equation}
\Delta_{\gamma_{p}}\W (p):= \mon_{\gamma_p} \W (p) =\int_{\gamma_p}\log y\, d\log x \,.
\label{Weff_monodromy}
\end{equation}
Clearly, $\gamma_p \cdot \W(p)$ and $\W(p)$ produce the same space of SUSY parameters \eqref{Vxiyi}. Let us note that the integral (\ref{Weff_monodromy}) in general depends on $p$ since $\log y$ may have a nontrivial monodromy along $\gamma_p$. It is easy to see that\footnote{The numbers $C_\gamma$ and $n_\gamma$ depend only on the homology class of the cycle $\gamma_p$. However the choice of the branch of $\log x$ in the r.h.s. of (\ref{Weff_monodromy_res}) (and thus the numerical result for $\Delta_{\gamma_{p}}\W(p)$ ) does depend on its homotopy class.}
\begin{equation}
 \Delta_{\gamma_{p}}\W (p)=C_\gamma+2\pi i n_\gamma \log x,\qquad \gamma\in H_1(\Cs)
 \label{Weff_monodromy_res}
\end{equation}
where $\gamma$ is the image of $\gamma_p$ under the Hurewicz map\footnote{There is also a version of
the Hurewicz homomorphism for pointed spaces. However, since we are mostly interested in path-connected $\CV$ and in choice of $\gamma_p$ up to isomorphism, to avoid clutter sometimes we write $\pi_1 (\CV) \cong \pi_1 (\CV,p)$.}
\begin{equation}
\pi_1(\CV) \; \stackrel{{h}}{\longrightarrow} \; H_1(\CV) \cong \pi_1(\CV) / [\pi_1(\CV), \pi_1(\CV)]
\end{equation}
and
\begin{equation}
 n_\gamma = \frac{\Delta_\gamma \log y}{2\pi i} \in \mathbb{Z} \,.
\end{equation}
Similarly one can define
\begin{equation}
 m_\gamma = \frac{\Delta_\gamma \log x}{2\pi i} \in \mathbb{Z} \,.
\label{magnetic_charge}
\end{equation}
Since $\W(p)$ and $\gamma_p\cdot\W(p)$ describe equivalent theories one can consider a one-dimensional defect $L_\gamma$ in the world volume of the 2d theory on $D \cong \mathbb{R}^2$ separating two regions with different effective twisted superpotential which can be realized by continuos deformation of the theory along the path $\gamma_x$ in the parameter moduli space $\Cs$. Consider the limit when the width of the defect tends to zero, so that the defect can be interpreted as a ``duality wall'' separating two equivalent theories in domains $D_1$ and $D_2$ (see Fig. \ref{figure_2d-defect}).

Replacing $\log x \to \Sigma_x$ in (\ref{Weff_monodromy_res}) it is easy to see that the Lagrangian of the theory on $D_2$ differs from the Lagrangian of the theory on $D_1$ by $2\pi i n_\gamma F_x$ where $F_x$ is the field strength for the $U(1)_x$ vector field. This means that the defect $L_\gamma$ changes coupling of the $U(1)_x$ gauge field $A_x$ of the 4d theory to the boundary by
\begin{equation}
2\pi i n_\gamma\int_{D_1} F_x \; = \; 2\pi i n_\gamma \int_{L_\gamma} A_x \,.
\label{XFAact}
\end{equation}
Therefore, the defect $L_\gamma$ can be considered as a particle with the charge $n_\gamma$ with respect to the symmetry group $U(1)_x$. By electric-magnetic duality $m_\gamma$ is its magnetic charge. As usual, the mass of the particle is given by $|\Delta_\gamma\W|$. Thus, the defects in question form a family of dyons labeled by the elements of $\pi_1(\CV)$ with charges given by the homomorphism $c_*$, which in turn is induced by the embedding
$c: \CV \hookrightarrow \C^* \times \C^*$,
\begin{equation}
\begin{array}{ccccl}
 \pi_1(\Cs) & \stackrel{{h}}{\longrightarrow} & H_1(\Cs) & \stackrel{{c_*}}{\longrightarrow} &
 \mathbb{Z}^2=H_1(\mathbb{C}^* \times \mathbb{C}^*)=:\Lambda_\text{em} \\
 \gamma_p & \longmapsto & \gamma & \longmapsto & (n_\gamma, m_\gamma) \,.
\end{array}
\label{charge_map}
\end{equation}
Note, the lattice $\Lambda_\text{em}$ introduced here is precisely the lattice of electric and magnetic charges in the 4d theory on half-space discussed around \eqref{4d_moduli} in which the symmetry $U(1)_x$ is gauged. (Again, we remind that generalization to theories with larger symmetry groups of rank $n>1$ is straightforward.) The image of $H_1 (\CV)$ under the map $c_*$ defines a sublattice of $\Lambda_{\text{em}}$. This is the lattice of charges carried by line operators that exist in 3d $\CN=2$ boundary theory $\CT$,
\beq
\Lambda_{3d} := c_* H_1 (\CV)
\label{3dlattice}
\eeq
The multiplication in $\pi_1(\Cs)$ corresponds to 3-particle interaction or, equivalently, to defect junctions in 3d $\CN=2$ theory $\CT$. In section~\ref{sect_branes} we shall explain how these dyons arise in the string theory picture.

The defect $L_\gamma$ can be interpreted as a 2d analogue of the four-dimensional Janus configuration \cite{Janus1,Janus2,Janus3} since as one goes across $L_\gamma$ the 2d effective theta angle $\Im\log y$ changes by $2\pi n_\gamma$. Instead of the Chern-Simons action on the world-volume of a Janus domain wall \cite{GW2}, the defects $L_{\gamma}$ considered here carry the standard action of a charged particle, {\it cf.} \eqref{XFAact}.

\subsection{Lift to 3d: tropical limit} \label{sect_tropical_limit}

In order to make connection between periods and line operators more explicit, it is convenient to introduce logarithmic variables
\beq
X = \log x
\qquad , \qquad
Y = \log y
\eeq
Of course, these variables are defined only up to integer multiples of $2 \pi i$. Indeed, since $\CV$ is defined by polynomial equations \eqref{Vxiyi}-\eqref{Acurve} in $x$ and $y$, they are invariant under the integer shifts
\beq
\text{Wilson} ~:~ Y \to Y + 2\pi i
\qquad
\Delta \tilde \CW = 2\pi i \log x
\label{Wilson}
\eeq
and
\beq
\text{'t Hooft} ~:~ X \to X + 2\pi i
\label{tHooft}
\eeq
which, according to the discussion in the previous section, we can identify with Wilson and 't Hooft line operators, respectively.

In general, the structure of the fundamental group $\pi_1 (\CV)$ can be rather complicated. This, in part, is what makes this story rich and interesting.
However, some information about its image \eqref{3dlattice} under the charge map can be read off directly from the ``shape'' of the curve $\CV$. In order to extract this information, it is convenient to look at the tropical limit of $\CV$ that physically corresponds to a decompactification of the circle,
\beq
S^1_t \times \R^2 \quad \leadsto \quad \R^3
\label{liftto3d}
\eeq
which will be explained in more detail later. In this limit, the supersymmetric parameters include only real parts of $X$ and $Y$, so that a complex algebraic curve \eqref{Acurve} turns into a real 1-dimensional graph $\CV_\trop$ in the $(\text{Re} \, X, \text{Re}\, Y)$ plane, that follows the skeleton of the amoeba of $\CV$, see Figure \ref{fig:curve2}. Asymptotic regions of this ``real moduli space'' of SUSY parameters are semi-infinite rays of the form
\beq
(\log |x|, \log |y|) = r (m,n)
\,, \qquad r \to + \infty \,,
\eeq
that correspond to tentacles of the amoeba. In other words, each semi-infinite ray or each tentacle of the amoeba is characterized by the charge vector $(n,m)$ that, without loss of generality, we can take to be a relatively prime pair of integers. These tentacles represent directions in the charge lattice along which the index $\CI (n,m)$ exhibits only ``linear growth'' (as opposed to generic, quadratic\footnote{More precisely, it refers to the rate of growth of the leading R-charge $R$ in the $q$-expansion of the index, $\CI (n,m) = a q^R + \ldots$, as $(n,m) \to \infty$.} growth) and signal presence of chiral multiplets of charge $(n,m)$ \cite{DGGindex}.

In the curve $\CV$ defined by a zero locus of the polynomial \eqref{Acurve}, each tentacle corresponds to an asymptotic region which has topology $\R_+ \times S^1$, in particular, it has a non-contractible cycle that corresponds to changing the value of the 2d background $\theta$-angle and the Wilson line of the background $U(1)_x$ gauge field:
\beq
\begin{array}{rcl}
\arg x & = & \oint_{S^1_t} A \\[.1cm]
\arg y & = & \theta
\end{array}
\eeq
Specifically, a tentacle labeled by the charge vector $(n,m)$ has a non-contractible cycle
\beq
(\arg x, \arg y) = (m \varphi, n \varphi)
\,, \qquad \varphi \in [0,2 \pi )
\eeq
It is clear that such cycles around cusps or asymptotic regions on the curve $\CV$ are non-contractible on all of $\CV$. Therefore, we obtain an important conclusion about the spectrum of charges of line operators:
\beq
\text{Span} \, ( \{ n,m \}_{\text{tentacles}} ) \; \subseteq \; \Lambda_{3d}
\label{spancharge}
\eeq
In other words, the spectrum of charges of line operators contains the span of all the charge vectors that determine directions of tentacles of the amoeba of $\CV$.
Note, that for curves $\CV$ of genus zero, {\it i.e.} for curves whose tropical limits are trees ({\it i.e.} graphs without closed loops), the relation \eqref{spancharge} becomes an equality.

Once we introduced $\Lambda_{\text{em}}$ and $\Lambda_{3d}$, we can define a quotient
\beq
\Lambda_{\text{em}} / \Lambda_{3d}
\eeq
that classifies flux sectors ({\it cf.} flux vacua \cite{GVW}) of the effective two-dimensional theory on $D$ modulo those connected by solitons (domain walls).

Let us consider the cycles that belong to the subgroup $E \subset H_1(\Cs)$ defined by the condition that the following sequence is exact:
\begin{equation}
0\rightarrow E\hookrightarrow H_1(\Cs) \stackrel{\trop}{\longrightarrow} H_1(\Cs_\trop)\rightarrow 0
\end{equation}
where ``$\trop$'' is the natural map which ``forgets'' how a cycle winds around the tubes that become lines in the tropical limit. The group $F := H_1(\Cs_\trop)\cong H_1(\Cs)/E$ is a free abelian group generated by the elements associated to finite faces of the graph $\Cs_\trop$:
\begin{equation}
 F=\langle \gamma_f\,|\,f\in \text{ faces of $\Cs_\trop$ }\rangle
\end{equation}
 The group $E$ is generated by its (oriented) edges:
\begin{equation}
 E=\langle \gamma_e\,|\,e\in \text{ edges of $\Cs_\trop$ }\rangle/
 \langle \pm\gamma_{e_1}\pm\gamma_{e_2}\pm\gamma_{e_3}\,|\,e_1,e_2,e_3 \text{ have common vertex } \rangle
\end{equation}
In particular, it contains all cycles \eqref{spancharge} associated to the tentacles. We can always embed $F\subset H_1(\Cs)$ by making a choice of how the cycles from $F$ pass along the tubes and pairs of pants associated with the vertices of the graph $\Cs_\trop$, {\it cf.} \cite{DMO}. At least when the curve is non-degenerate it is possible to choose $F\subset \ker{c_*}$, where ${c_*}$ is defined in (\ref{charge_map}). Even though $\CV$ is singular in a number of examples considered below, for the sake of simplicity we pretend that it is not the case and
\begin{equation}
 H_1(\Cs)=E \oplus F \,.
\end{equation}

Now let us now consider the dependence of the curve $\Cs$ on the complex structure parameters $\{t_i\}$. Suppose that the parameters are chosen so that $\{\log t_i\}$ are flat coordinates on the moduli space of the curve $\CV$. They correspond to the Coulomb and mass parameters of the 5d gauge theory engineered by the corresponding toric Calabi-Yau 3-fold \cite{engineering,Nikita,NLawrence,pqwebs}. Then, for any contour $\gamma_e \in E \subset H_1(\Cs)$ associated to the edge of the web $\Cs_\trop$ we have
\begin{equation}
 \Delta_{\gamma_e}\W (x):=\int_{\gamma_e}\log y\, d\log x
 =2\pi^2c_{\gamma_e}+2\pi i\sum_{i}q_{\gamma_e,i}\log t_i+2\pi i n_{\gamma_e} \log x,\qquad
 c_{\gamma_e},\, q_{\gamma_e,i},\, n_{\gamma_e}  \in\mathbb{Z}
 \label{periods_moduli}
\end{equation}
If we interpret $\log t_i$ as a v.e.v. of a 2d background twisted chiral field $\Sigma_{t_i}$, then one can deduce that the defect labeled by $\gamma$ carries charges $q_{\gamma,i}$ with respect to the corresponding global symmetries $U(1)_{t_i}$. All these line defects in 2d can be naturally lifted to 1d objects in three dimensions, in agreement with their description as string endpoints in brane constructions of 3d $\CN=2$ theories (see section \ref{sect_branes} for details).

The non-constant part of (\ref{periods_moduli}) can be easily read off from the tropical geometry where $\log t_i$ are simply length parameters. If we parameterize the corresponding edge as
\begin{equation}
e \subset \{(m \xi+\alpha,n\xi+\beta)\;|\;\xi\in \mathbb{R}\}\subset \{(\Re X,\Re Y)\}\cong\mathbb{R}^2
\end{equation}
so that $\alpha$ and $\beta$ are linear combinations of $\log t_i$,
then the corresponding electric and magnetic charges are given by the orientation of the edge:
\begin{equation}
 (m_{\gamma_e},n_{\gamma_e}):={c_*}\gamma_e=(m,n)
\end{equation}
and
\begin{equation}
 \Delta_{\gamma_e}\W (x):=\int_{\gamma_e}\log y\, d\log x
 =2\pi^2 c_{\gamma_e}+2\pi i(m \alpha - n \beta)+2\pi i n_{\gamma_e} \log x,\qquad c_{\gamma_e},\,q_{\gamma_e,i},\,n_{\gamma_e}\in\mathbb{Z}
 \label{DW_geom_res}
\end{equation}

For a cycle $\gamma_f\in F\subset H_1(\Cs)$ associated to a face $f$ of the web $\Cs_\trop$ the situation is slightly more complicated. In general, the period is expected to have the following form\footnote{It follows, for example, from the fact that the prepotential for the curve $\CV$, as a function of Coulomb and mass parameters $a_i\sim \log t_i$ has the following form:
\begin{equation}
\CF(a)=P_3(a)+\sum_{\beta}n_\beta\text{Li}_3(e^{-\beta\cdot a})
\end{equation}
where $P_3(a)$ is a cubic polynomial and $n_\beta$ are (integer) genus zero Gopakumar-Vafa invariants \cite{GV}. Then (\ref{periods_faces}) follows as its derivative.
}:
\begin{equation}
 \Delta_{\gamma_f}\W (x):=\int_{\gamma_f}\log y\, d\log x
 =\sum_{i,j}\frac{n_{\gamma_f,i,j}}{2}\log t_i\log t_j+2\pi i\sum_{i}q_{\gamma_f,i}\log t_i+\sum_{\alpha}N_{\gamma_f,\alpha} \dilog(\prod_i t_i^{-\alpha_i})
 \label{periods_faces}
\end{equation}
The quadratic part calculates the area of the face $f$ of the tropical curve. Let us remind that we have chosen $F$ so that ${c_*}|_F=0$, where $c_*$ is the charge map introduced in \eqref{charge_map}.

However, as will be shown in section \ref{sect_examples}, for any theory originating from some Lagrangian description (under certain mild assumptions)  all these periods are actually trivial, that is
\begin{equation}
\begin{array}{lr}
 n_{i,j} &=0 \,, \\
 N_{\alpha}&=0 \,.
\end{array}
\label{periods_faces_cond}
\end{equation}
This is related to the conjecture of Aganagic and Vafa \cite{AV} (since every knot $K$ is supposed to correspond \cite{DGH,DGG1,FGS} to a theory $\CT_{K}$ with a Lagrangian description).

In general, the quadratic terms in (\ref{periods_faces}) indicate that these defects should be lifted to 2d objects in 3d described by $\mathcal{N}=(0,2)$ theories. The 2d nature of these objects is also predicted by the string theory picture which will be considered in section \ref{sect_branes}.

\subsection{``Small'' line operators }

Consider the ``half-index''  $\mathcal{I}_{S^1_t\times_q D}(\CT)$ of the theory $\CT$ on $S^1_t \times_q D$. It has the following asymptotic behavior, {\it cf.} \eqref{ZZZasymp}:
\begin{equation}
\mathcal{I}_{S^1_t\times_q D}(\CT) \; \simeq \; e^{\frac{\W}{\hbar}} \,,\qquad \hbar\rightarrow 0 \,,
\end{equation}
where $\hbar := \log q$ defines the conformal structure of the boundary torus $S^1_t \times_q S^1_{sp} = \partial (S^1_t \times_q D)$. Since $\hbar$ is basically the ratio of the radii of the ``temporal'' and ``spatial'' circles of the boundary torus, in this limit we can treat the former as ``small'' and the latter as ``large,''
\begin{equation}
\begin{array}{lr}
S^1_t : & \text{``small'' as}~\hbar\rightarrow 0 \,, \\
S^1_{sp} : & \text{``large'' as}~\hbar\rightarrow 0 \,.
\end{array}
\end{equation}
The insertion of Wilson and 't Hooft operators (with respect to $U(1)_x$) wrapping the ``temporal'' (or, what we now call ``small'') circle corresponds to multiplication by $x$ and $y$, respectively. Thus, the Wilson operator supported on $S^1_t$ is equivalent to the ``small'' shift of the effective twisted superpotential:
\begin{equation}
\Delta\W =\hbar \log x
\end{equation}
as opposed to the ``large'' shift (\ref{Weff_monodromy_res}) associated to a Wilson operator \eqref{Wilson} supported on $S^1_{sp}$.
Similarly, the shift of the effective twisted superpotential associated to the ``small'' 't Hooft loop,
\begin{equation}
\Delta\W \; = \; \hbar \log y \,,
\end{equation}
can be achieved by shifting $\log x$ by the amount
\begin{equation}
\Delta \log x \; = \; \hbar \,,
\end{equation}
which is again ``small'' compared to the ``large'' shift (\ref{magnetic_charge}) associated to the 't Hooft operator \eqref{tHooft} supported on $S^1_{sp}$.

Upon compactification on a ``temporal'' circle $S^1_t$, one finds the effective 2d $\CN=(2,2)$ theory, which is the home to the twisted superpotential $\CW$ as well as the curve $\CV$. From the viewpoint of this 2d theory, the ``large'' and ``small'' line operators have rather different nature. The former are one-dimensional object in this 2d theory, whereas the latter are local operators. Upon the lift \eqref{liftto3d} to three dimensions ``small'' operators become 1d objects and ``large'' operators become 1d or 2d objects.


\subsection{Parameter walls}

\begin{figure}
\center{\includegraphics[scale=1.2]{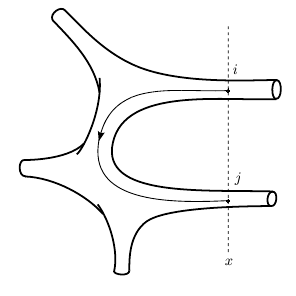}}
\caption{\label{domainwallperiod}Period corresponding to domain wall interpolating between vacuum $i$ and vacuum $j$ of the theory at a given value of the SUSY parameter $x$.}
\end{figure}

In addition to the periods that correspond to line operators, the other two types of periods shown in figure~\ref{fig:curve2} have the following interpretation. The periods on closed cycles correspond to a sequence of domain walls that interpolate between the same vacua of the theory. On the other hand, the periods that start from one asymptotic region  and go to another asymptotic region correspond to domain walls that interpolate between two infinitely massive vacua of the same theory possibly at different values of the SUSY parameters. Although natural from the point of view of homology, these periods have  an awkward interpretation in the gauge theory. Instead it is more natural to consider periods on cycles that interpolate from a point ``$i$" on one tentacle to a point ``$j$" on another tentacle at the same value of SUSY parameters, see figure~\ref{domainwallperiod}. They correspond to domain walls interpolating between two different vacua. We discuss these objects in what follows.

The tension of the domain walls is $|\Delta{\W}_{ij}|=|{\W}_{i}-{\W}_{j}|$ which is computed by the period integral along the path shown in figure~\ref{domainwallperiod}. The spectrum of such walls (sometimes also called kinks) is interesting in its own right. The whole idea of \cite{FMVW,FLMW,LercheW,CVclassification} is that one can identify $\CN=2$ superconformal field theory in two dimensions by the soliton spectrum of its massive deformations. It was shown in \cite{CVclassification} that the soliton spectrum captures the R-charge of the Ramond ground states \footnote{This theme of relating the BPS spectrum to the superconformal spectrum has been pursued recently for three, four and five dimensional supersymmetric theories \cite{Iqbal:2012xm}. }. A more precise statement is: Let $N_{ij}$ be the number (with sign) of kinks interpolating between vacuum $i$ and vacuum
$j$. Define the ordered product,
\begin{equation}
S=\prod_{ij}^{\curvearrowright}M_{ij},\qquad M_{ij}:= I-N_{ij}T_{ij}.
\end{equation}
Here, the product has been taken over the kinks and ordered according to the phase of their central charge $\Delta{\W}$. $I$ is the identity matrix and $T_{ij}$ is the matrix whose only $ij$-th entry is $1$ and others $0$. Then,
\begin{equation}
\mbox{Tr}(SS^{-T})^{k}=\sum_{j}e^{2\pi iR_{j}k}\label{kink-conformal}
\end{equation}
where $R_{j}$ on the right hand side of the equation are the $R$ charges of the Ramond ground states. In fact, the RHS is a special limit of the elliptic genus $Z(y,e^{2\pi i\tau})|_{y=e^{2\pi ik}}$, a quantity clearly associated with the superconformal theory. Interestingly, in this limit, the elliptic genus turns out to be independent of modulus. 

This intimate relation between the BPS spectrum and the superconformal spectrum was later understood in \cite{Cecotti:2010qn} by formulating a partition function $\mbox{Tr}(-1)^{F}e^{2\pi ikJ}$ of the two dimensional theory on a torus and evaluating it in two ways. At high energies, the partition function is clearly the previously mentioned limit of the elliptic genus which gives the RHS of the eq. (\ref{kink-conformal}). At low energies, it is natural to evaluate the path integral in canonical quantization. Because of the insertion of R-twist $e^{2\pi i k J}$, this quantum mechanics turns out to be time dependent. The transition amplitude is dominated by a sequence of kink configurations, each fundamental kink contributing $M_{ij}$. The time ordering of the kinks follows from the phase ordering of their central charge $\Delta{\W}_{ij}$. The transition amplitude on $1/k$-th slice of the cylinder is $SS^{-T}$. Gluing $k$ of such cylinders and taking the trace gives us the LHS of eq. (\ref{kink-conformal}). 

\begin{figure}
\begin{centering}
\includegraphics[scale=0.4]{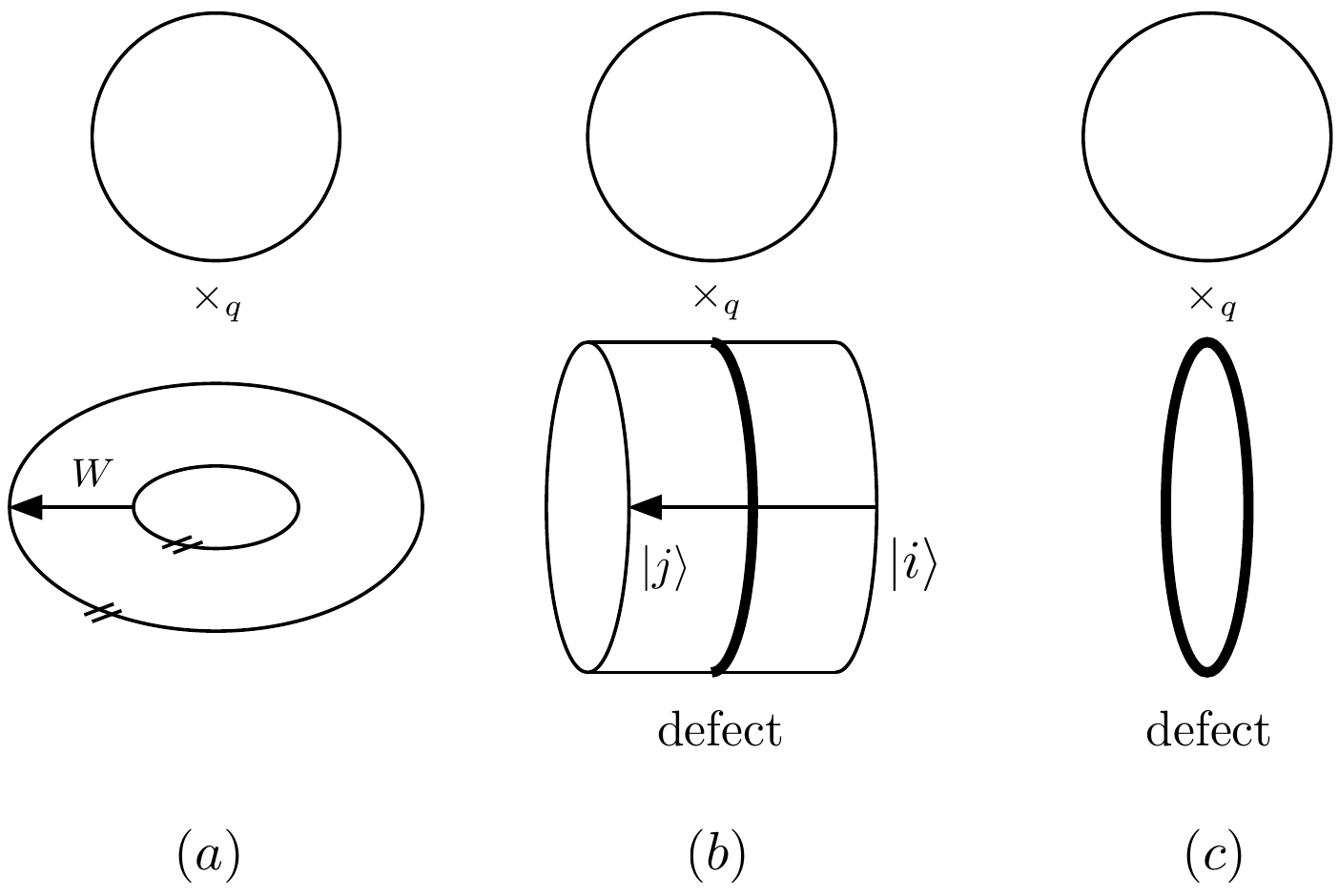}
\par\end{centering}
\caption{In all the figures $\times_{q}$ stands for the fibration on $S^{1}$ with equivariant parameter $q$. Figure $(a)$ shows the space on which we define $Z_{{\cal T}}^{T^{3}}(q,k)$. The line labelled $W$ stands for the twist operator insertion $e^{2\pi ikJ}$. Unwrapping the torus produces a cylindrical slice with vacuum $|i\rangle$ on one side and vacuum $|j\rangle$ on the other. The path integral on this geometry is saturated by the BPS kink configuration. This is shown in figure $(b)$. Figure $(c)$ isolates the support of the defect and illustrates that it is a two-torus with nome $q$.}
\label{domainwall-geometry}
\end{figure}

In the present context, a similar computation can be performed. The partition function relevant to derive such relation is formulated by considering the three dimensional ${\cal N}=2$ theory ${\cal T}$ on $T^{3}$ with an insertion of the twist operator $e^{2\pi ikJ}$, as shown in figure \ref{domainwall-geometry}. Let this partition function be $Z_{{\cal T}}^{T^{3}}(q,k)$. Note that equivariant parameter $q$ considered before in the geometry $S^{1}\times_{q}D$ maps to one of the modular parameters of $T^{3}$. The kink solutions are lifted to two dimensional half-BPS domain wall solutions in theory ${\cal T}$. This defect preserves $(0,2)$ supersymmetry in the 2d support of the defect. Moreover, this support is precisely a $T^{2}$ with modular parameter $q$, as explained in figure \ref{domainwall-geometry}.
We can evaluate $Z_{{\cal T}}^{T^{3}}$ at low energies, where the path integral will again be dominated by the domain wall defects. Instead of simply contributing $S_{ij}$ as in eq. \eqref{kink-conformal}, the wall contributes its full partition function. As the support of the domain wall is a torus with modular parameter $q$, this partition function is simply the elliptic genus $Z_{ij}(q)$ of the $(0,2)$ theory living on the defect. This leads us to conjecture the following relation between the $T^{3}$ partition function of the 3d ${\cal N}=2$ superconformal theory and the spectrum of its domain walls:
\begin{equation}
\mbox{Tr}(ZZ^{-T})^{k}=Z_{{\cal T}}^{T^{3}}(q,k)|_{SCFT},
\end{equation}
where $Z$ is the appropriately phase ordered matrix product constructed out of the elliptic genera $Z_{ij}$ of the kink interpolating from vacuum $i$ to vacuum $j$.

\section{Transformation walls and 2d $(0,2)$ theories} \label{sect_transf_walls}

There are other half-BPS co-dimension one defects that are present in a 3d $\CN=2$ superconformal theory $\CT$. In this section we consider such 3d theory with $U(1)$ global symmetry and discuss the defect wall associated to the $SL(2,{\mathbb Z})$ transformation. This transformation was defined by Witten in \cite{Witten:2003ya}\footnote{One can also consider a 3d theory with $U(1)^N$ global symmetry and the action of $Sp(2N,{\mathbb Z})$ on it.}. Our main interest will be in identifying the 2d degrees of freedom living on the  defect. Before getting into this analysis let us describe the geometric setup and associated partition functions.

For our purposes, it is most convenient to place $\CT$ on $S^1 \times_q D$. The twisted product $\times_q$ can be thought of as turning on a Wilson line on $S^1$ for the $SO(2)$ rotational symmetry of $D$. We take the defect to live on the boundary torus. The parameter $q$ finds a natural interpretation on the boundary, it is the amount with which the boundary circle of $D$ is rotated as one goes around $S^1$. In other words $q:= e^{2\pi i \tau}$ is the nome of the boundary $T^2$ and $\tau$, the modular parameter. The partition function of the bulk theory in this background is called the K-theoretic vortex partition function or the holomorphic block and has been studied in \cite{Dimofte:2010tz,BDP}. The half-BPS defect living on the boundary defines a 2d $(0,2)$ superconformal theory on $T^2$ with nome $q$. We will be interested in the partition function of the bulk-boundary system. The partition function of the boundary theory is nothing but the elliptic genus.

The 3d ${\cal N}=2$ superconformal theory has four complex supercharges ${\cal Q}_{\alpha}^{i}$, here $i=1,2$ is the $SO(2)$ R-symmetry index and $\alpha=1,2$ is $SU(2)$ rotation index. It also has four conformal supercharges ${\cal S}_{i}^{\alpha}=({\cal Q}_{\alpha}^{i})^{\dagger}$. Out of the four supercharges, the 2d boundary or the defect preserves only two supercharges, ${\cal Q}_{+}^{1}$ and ${\cal Q}_{+}^{2}$. In two dimensions they generate $(0,2)$ supersymmetry \footnote{The half-BPS wall defect can either preserve $(0,2)$ or $(1,1)$ supersymmetry. We have focused on the former case. See section \ref{IIBstring} for an argument about $(0,2)$ supersymmetry for parameter walls.}. This supersymmetry is enhanced to $(0,2)$ superconformal symmetry in the infrared. The most interesting commutation relations of the algebra (in the NS sector) are
\begin{equation}
\{{\cal Q}_{+}^{1,2},({\cal Q}_{+}^{1,2})^{\dagger}\}=H_{R}\mp\frac{1}{2}J_{R}
\end{equation}
where $H_{R}$ is the dilation generator in the right moving sector and $J_{R}$ is the $SO(2)$ R symmetry generator. Note that the R symmetry is only present in the right moving sector.

\subsection{The index of 2d $(0,2)$ theories}

The elliptic genus of the boundary theory can be computed as the index in the radial quantization:
\begin{equation}
{\cal I}:=\mbox{Tr}(-1)^{F}q^{H_{L}}a^{f}.
\end{equation}
Here $H_{L}$ is the left-moving dilation generator and $f$ is a generator of the $U(1)$ flavor symmetry\footnote{In conventional definition of the elliptic genus, $a$ is set to $1$. So what we are considering here is really a \emph{flavored} elliptic genus}. This trace is protected because both $H_{L}$ and $f$ commute with the supercharges. Because the index is independent of the coupling, we can simplify our computations by taking the free field limit and considering each multiplet separately. Most of the $(0,2)$ symmetric Lagrangians can be constructed out of three types of multiplets: chiral multiplet, Fermi multiplet and vector multiplet. Below we compute the contribution of each of the three types of multiplets to ${\cal I}$.

\subsection*{Chiral multiplet}

The 2d $(0,2)$ chiral multiplet $\Phi$ satisfies $D_{+}^{2}\Phi=0$ and has the superspace expansion:
\begin{equation}
\Phi=\phi+\theta_{1}^{+}\psi_{+}-i\theta_{1}^{+}\theta_{2}^{+}\partial_{+}\phi.
\end{equation}
Let us take this chiral field to have charge $1$ under some flavor
symmetry $f$. The boson $\phi$ and all its descendants $\partial_{-}^{n}\phi$
along with their conjugates contribute
\begin{equation}
\left(\prod_{i=0}(1-xq^{i})(1-x^{-1}q^{i})\right)^{-1}.
\end{equation}
here, $x$ is a fugacity for the $U(1)$ flavor symmetry $f$. For the fermion $\psi_{+}$, all its modes and their conjugate contribute. Also, the equation motion $\partial_{-}\psi_{+}$ and all its descendants along with their conjugates contribute but with the opposite sign. As a result, only the contribution of the zero mode survives
\begin{equation}
(x^{\frac{1}{2}}-x^{-\frac{1}{2}}).
\end{equation}
Combining,
\begin{equation}
{\cal I}_{\Phi}(x;q)=\frac{(x^{\frac{1}{2}}-x^{-\frac{1}{2}})}{\prod_{i=0}^{\infty}(1-xq^{i})(1-x^{-1}q^{i})}=: x^{\frac{1}{2}}\theta(x;q)^{-1}.
\end{equation}
Here we have assumed the canonical left-moving dimension for the scalar field $H_{L}=0$. The index of the chiral multiplet with a general dimension $H_{L}=h_{L}$ is ${\cal I}_{\Phi}(xq^{h_{L}};q)$.

\subsection*{Fermi multiplet}

The Fermi multiplet $\Psi$ also satisfies $D_{+}^{2}\Psi=0$. As the name suggests, it has a fermionic primary. Its expansion in the superspace is:
\begin{equation}
\Psi=\psi_{-}+\theta_{1}^{+}G-i\theta_{1}^{+}\theta_{2}^{+}\partial_{+}\psi_{-}.
\end{equation}
Only the letter $\psi_{-}$ and its conjugate contribute to the index, including their zero modes. The index of the fermi multiplet is
\begin{equation}
{\cal I}_{\Psi}^{(R)}(x;q)=(x^{\frac{1}{2}}-x^{-\frac{1}{2}})\prod_{i=1}^{\infty}(1-q^{i}x)(1-q^{i}x^{-1})=-x^{-\frac{1}{2}}\theta(x;q).
\end{equation}
again, $x$ is a global $U(1)$ symmetry fugacity. Here also we have assumed $H_{L}=0$, for a Fermi multiplet with general dimension $H_{L}=h_{L}$, the index is ${\cal I}_{\Psi}(xq^{h_{L}};q)$.

\subsection*{Vector multiplet}

The easiest way of determining the index of the vector multiplet is by using the super-Higgs mechanism. As the gauge symmetry is spontaneously broken, the vector multiplet eats a massless chiral multiplet to become massive. The index formulation of this phenomenon is
\begin{equation}
{\cal I}_{V}\mbox{Res}_{x\to1}{\cal I}_{\Phi}=1 \Rightarrow {\cal I}_{V}^{(R)}=(q;q)^{2}.
\end{equation}

We can put the indices of all the constituent multiplets together to compute the index of 2d $(0,2)$ theories. The simplest example would be that of a Fermi multiplet interacting with a chiral multiplet with a superpotential like coupling
\begin{equation}
S_{{\cal V}}=\int d^{2}xd\theta^{+}\Psi\Phi=\int d^{2}x(\phi G+\psi_{+}\psi_{-})
\end{equation}
The index of this theory is simply the product of the index of the chiral multiplet and the index of the Fermi multiplet. The global symmetry acts oppositely on these multiplets.
\begin{equation}
{\cal I}=-\frac{x^{-\frac{1}{2}}\theta(x;q)}{x^{\frac{1}{2}}\theta(x^{-1};q)}=1.
\end{equation}
Indeed, the coupling $S_{{\cal V}}$ generates the mass term for both bose and fermi degrees of freedom and in the infrared we get an empty theory. The index of the empty theory is expected to be $1$. Let us  remark that from the index of the basic multiplets one can also compute the index of the gauge theories that flow to superconformal fixed points. We simply multiply the index contribution of all multiplets and integrate over the gauge fugacity to impose the Gauss law. We will not be considering the gauge theory index any further.

\subsection{Duality wall in four dimensions}

\begin{figure}
\begin{centering}
\includegraphics[scale=0.3]{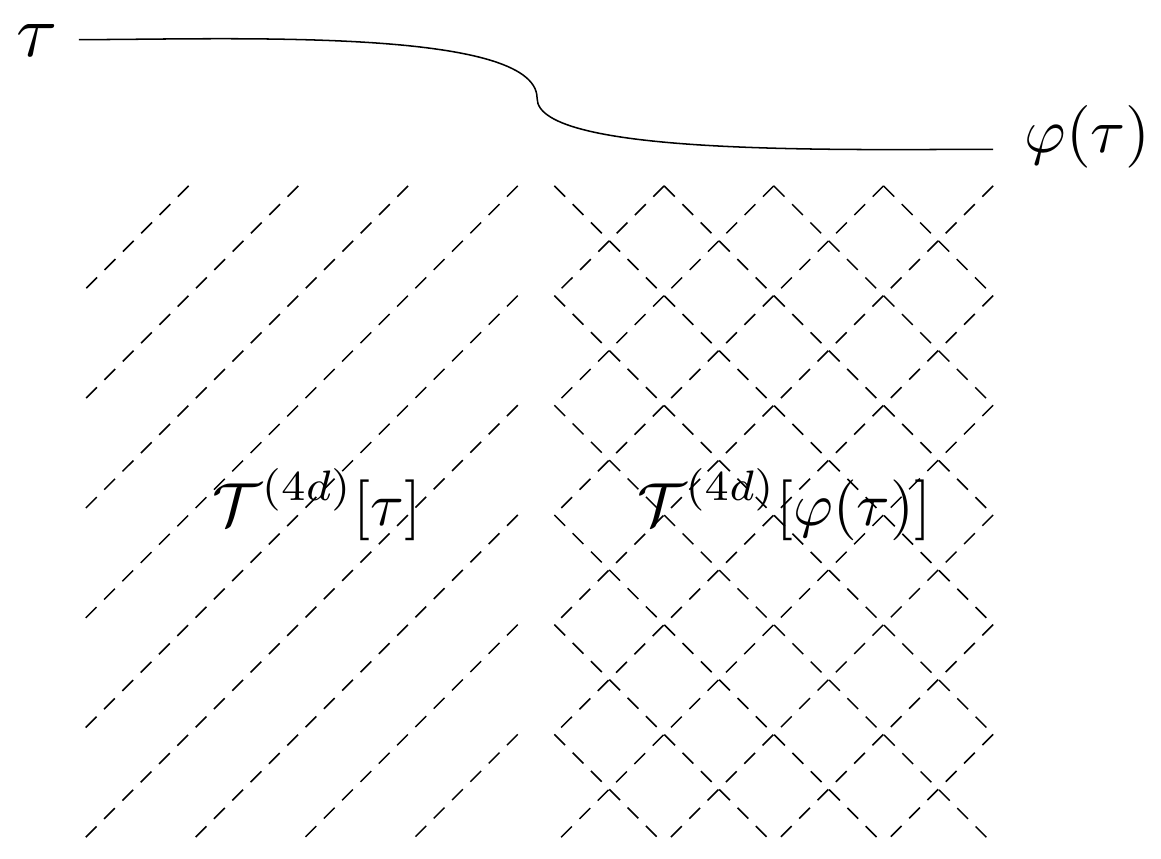}$\qquad$\includegraphics[scale=0.3]{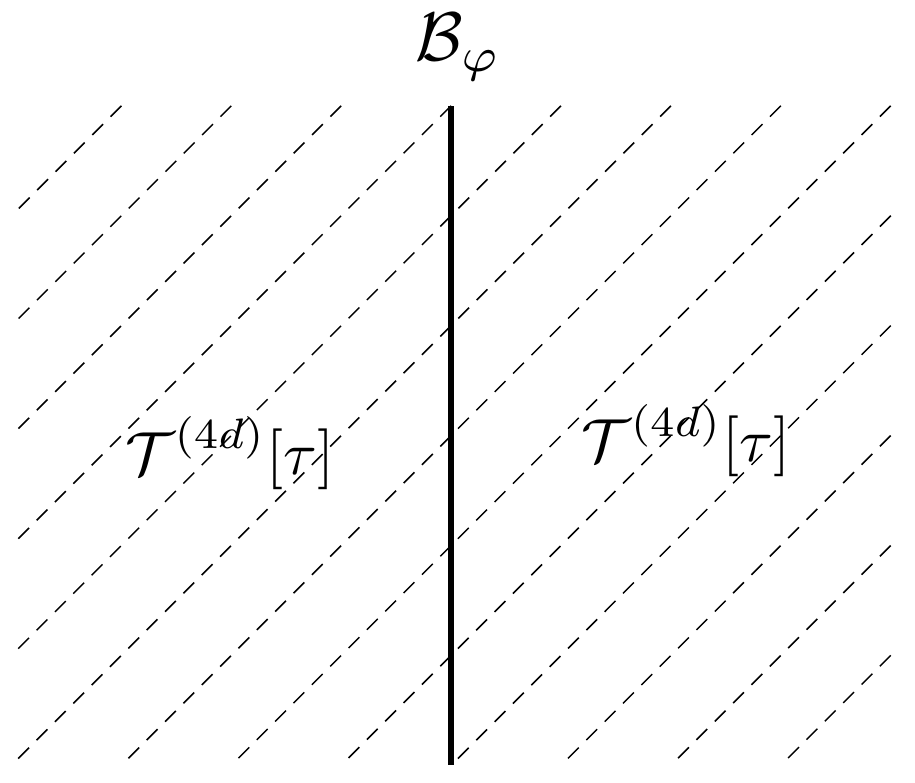}
\par\end{centering}
\caption{\label{wallsin4d}The first figure  shows the theory ${\cal T}^{(4d)}[\tau]$  on the left half-space and the dual theory ${\cal T}^{(4d)}[\varphi(\tau)]$ on the right half-space. In the second figure we have dualized the theory  on the right  back to ${\cal T}^{(4d)}[\tau]$ while introducing new degrees of freedom ${\cal B}_{\varphi}$ on the interface. }
\end{figure}

Let us get back to the subject of $SL(2,{\mathbb Z})$ transformation walls in three dimensional theories. A best way of understanding the  action of $SL(2,{\mathbb Z})$ on an $\CN=2$ 3d theory $\CT$ with $U(1)$ global symmetry is by coupling it to the 4d $\CN=4$  $U(1)$ gauge theory $\CT^{(4d)}$. Our strategy for studying the transformation wall in 3d is by thinking of it as a duality wall for a $3d-4d$ coupled system. It is convenient to first review the duality wall only in the 4d theory $\CT^{(4d)}$ theory (without boundary). The ${\cal N}=2$ $U(1)$ gauge theory ${\cal T}^{(4d)}[\tau]$ in four dimensions with coupling $\tau$ admits the action of $\varphi\in SL(2,\mathbb{Z})$ duality group
\begin{equation}
{\cal T}^{(4d)}[\tau]\sim{\cal T}^{(4d)}[\varphi(\tau)].
\end{equation}
Consider the setup in Fig.~\ref{wallsin4d}, where the coupling of the theory ${\cal T}$ changes from $\tau$ at $x_{2}=-\infty$ to $\varphi(\tau)$ at $x_{2}=+\infty$. We dualize theory just on the half space $x_{2}>0$ so that it has coupling $\tau$. In the process we have introduced new degrees of freedom ${\CB}_{\varphi}$ supported at the three dimensional interface $x_{2}=0$. As this co-dimension one defect or wall is associated to an element of the duality group, it also called a duality wall. The duality group $SL(2,Z)$ is generated by $T:\tau\to\tau+1$ and $S:\tau\to-1/\tau$. The three dimensional theory ${\CB}_{\varphi}$ can be explicitly identified for both generators. The theory ${\CB}_{\varphi=T}$ is an ${\cal N}=2$ Chern-Simons theory $AdA$ at unit level. Here $A$ is the four dimensional gauge field restricted to the wall. The theory ${\cal B}_{\varphi=S}$ has an ${\cal N}=2$ cross CS term $A_{L}dA_{R}$ at unit level, where $A_{L}$($A_{R}$) is the bulk gauge field on the left (right) side of the wall restricted to the defect.

\subsubsection{Duality wall with boundary}

Now let us consider the duality wall in $\CT^{(4d)}$  living on the half-space $x_{3}\geq0$ with half-BPS boundary conditions at $x_{3}=0$. The half-BPS boundary condition is defined by coupling ${\cal T}^{(4d)}$ to a general three dimensional ${\cal N}=2$ theory ${\cal T}$ with $U(1)$ flavor symmetry. This involves coupling the flavor current of ${\cal T}$ to the gauge field of ${\cal T}^{(4d)}$: $\int A^{(4d)}\wedge*J^{(3d)}$. Let us first consider the action of the duality group $SL(2,{\mathbb Z})$ on the bulk boundary system $({\cal T}^{(4d)},{\cal T})$ in the absence of any wall at $x_{2}=0$.

As before, the theory ${\cal T}^{(4d)}$ enjoys the $SL(2,{\mathbb Z})$ duality $\varphi:\tau\to\varphi(\tau)$, where $\tau$ is the gauge coupling and $\varphi$ is an element of $SL(2,{\mathbb Z})$. Moreover, the action of $\varphi$ on ${\cal T}$ can be defined by imposing the equivalence between the bulk-boundary tuples
\begin{equation}
({\cal T}[\tau]^{(4d)},{\cal T})\sim({\cal T}^{(4d)}[\varphi(\tau)],{\cal T}_{\varphi}).
\end{equation}
In \cite{Witten:2003ya} Witten explicitly identified the $SL(2,{\mathbb Z})$ action on the three dimensional theory ${\cal T}$. The element $T:\tau\to\tau+1$ change the $\theta$ angle by $2\pi$. The addition of $F\wedge F$ term to the four dimensional Lagrangian gives rise to a level $1$ Chern-Simons term $AdA$ on the boundary. This can be thought of as the addition of a background $\text{CS}_{1}$ term for the $U(1)$ flavor symmetry. Although more involved, it can be shown that the action of the element $S:\tau\to-1/\tau$ on ${\cal T}$ results in replacing the bulk-boundary interaction by $\int A^{(3d)}\wedge*J^{(3d)}+A^{(4d)}dA^{(3d)}$ where $A^{(3d)}$ is a new three dimensional field that gauges the $U(1)$ flavor symmetry. This is equivalent to coupling the four dimensional gauge field not to the flavor current but rather to the topological current associated to the gauging of the flavor current. To summarize,
\begin{eqnarray*}
T & : & \mbox{add background \ensuremath{\text{CS}_{1}} term for the \ensuremath{U(1)} flavor symmetry}\\
S & : & \mbox{replace the flavor symmetry by topological \ensuremath{U(1)} symmetry}
\end{eqnarray*}

\begin{figure}
\begin{centering}
\includegraphics[scale=0.3]{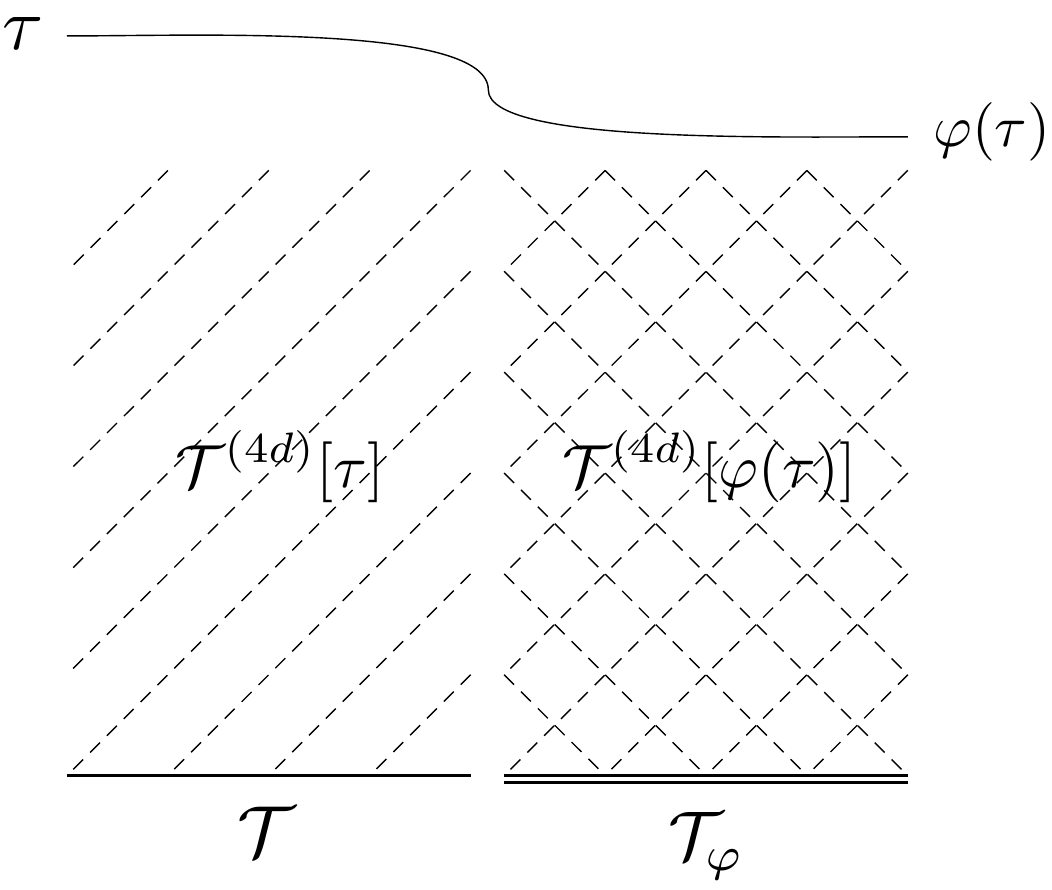}$\qquad$\includegraphics[scale=0.3]{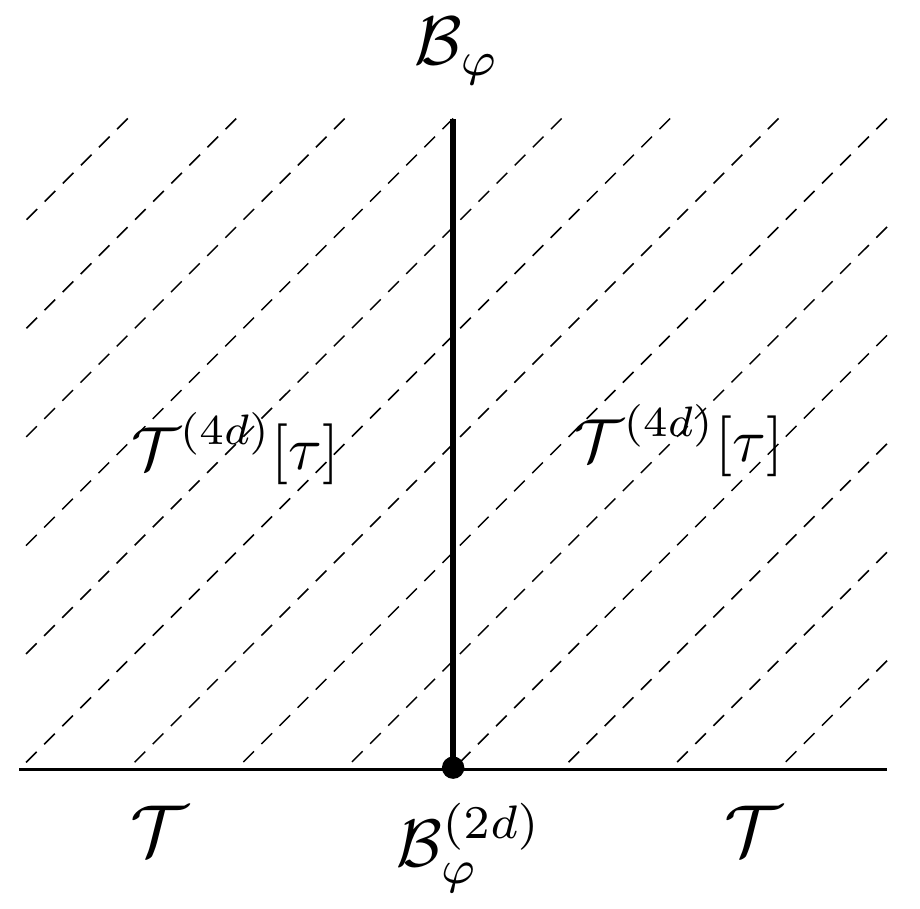}
\par\end{centering}
\caption{\label{wallwithboundary}The first figure  shows the system $({\cal T}^{(4d)}[\tau],{\cal T})$  on the left half-space and the dual system $({\cal T}^{(4d)}[\varphi(\tau)],{\cal T}_{\varphi})$ on the right half-space. In the second figure we have dualized the theory on the right while introducing new degrees of freedom ${\cal B}_{\varphi}$ on the interface. Now this interface itself has a two dimensional boundary.}
\end{figure}

\begin{figure}
\begin{centering}
\includegraphics[scale=0.4]{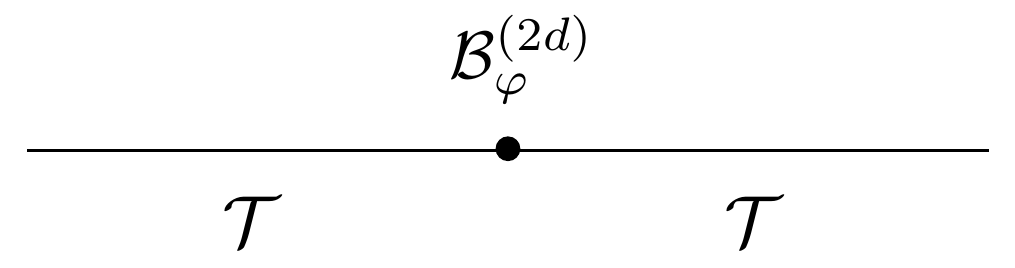}
\par\end{centering}
\caption{\label{2dwall}Decoupling the bulk from the boundary, we get only the three dimensional theory ${\cal T}$ with a duality transformation wall ${\cal B}^{(2d)}_{\varphi}$.}
\end{figure}

Now we are in the position of considering duality walls in this bulk-boundary system. Previously we considered the system with ${\cal T}^{(4d)}[\tau]$ for $x_{2}<0$ and the dual theory ${\cal T}^{(4d)}[\varphi(\tau)]$ for $x_{2}>0$. In the presence of boundary, we place the tuple $({\cal T}^{(4d)}[\tau],{\cal T})$ on the half-space $x_{2}<0$ and the ``dual'' tuple $({\cal T}^{(4d)}[\varphi(\tau)],{\cal T}_{\varphi})$ on the other half $x_{2}>0$. We dualize the bulk-boundary system on the later half back to $({\cal T}^{(4d)}[\tau],{\cal T})$ and in the process introduce new degrees of freedom ${\cal B}_{\varphi}$, just like before, but now living on the three dimensional \emph{half-space} $x_{3}>0$. Everywhere on the boundary, we have the theory ${\cal T}$ except at $x_{2}=0$ (and $x_{3}=0$). We have introduced new degrees of freedom on the two dimensional interface at $x_{2},x_{3}=0$ which are just the boundary degrees of freedom of ${\cal B}_{\varphi}$. This discussion is summarized in Fig. \ref{wallwithboundary}. Completely decoupling the four dimensional bulk theory living on the half-space $x_{3}>0$ gives us the definition of a two dimensional ``transformation wall'' for the three dimensional theory ${\cal T}$, see Fig \ref {2dwall}.

\subsection{Transformation wall in three dimensions} \label{sect_transf_wall_3d}

As noted earlier, for $\varphi=T$ the theory ${\cal B}_{\varphi}$ is an ${\cal N}=2$ Chern-Simons theory at unit level. Gauge invariance of the ${\cal N}=2$ CS action induces a unique 2d boundary theory ${\cal B}_{\varphi}^{(2d)}$: a $(0,2)$ supersymmetric $U(1)$ WZW model at level $1$. The index of this theory is just the index of a chiral multiplet. This has been evaluated in the previous section. We have
\begin{equation}
{\cal I}_{{\cal B}_{\varphi=T}^{(2d)}}(x)=x^{\frac{1}{2}}\theta(x;q)^{-1}
\end{equation}
This matches  (up to the factor $x^{\frac{1}{2}}$) with the observation in \cite{BDP} that in order to raise the CS level in theory ${\cal T}$ by $1$ on only the half space, one needs to insert the factor of $\theta(x;q)^{-1}$ in the $S^{1}\times_q D$ partition function of ${\cal T}$:
\begin{equation}
{\cal I}_{S^{1}\times_q D}^{{\cal T}}(x)\stackrel{T}{\longrightarrow}{\cal I}_{{\cal B}_{\varphi=T}^{(2d)}}(x){\cal I}_{S^{1}\times_q D}^{{\cal T}}(x).
\label{T-transf-Z}
\end{equation}
Since
\begin{equation}
 \theta(x;q)\approx e^{-\frac{1}{2\hbar}(\log x)^2},\qquad \log q= \hbar \rightarrow 0,
\end{equation}
in the limit $q\rightarrow 1$ (\ref{T-transf-Z}) is reduced to the correct $T$-transformation of the effective twisted superpotential and the functions $x,y$ on the curve:
\begin{equation}\label{T-transf-W}
 \W(x)\stackrel{T}{\longrightarrow} \W(x)+\frac{1}{2}(\log x)^2,
\end{equation}
\begin{equation}
 x\longrightarrow x,
\end{equation}
\begin{equation}
 y\longrightarrow yx.
\end{equation}

For $\varphi=S$, the theory ${\cal B}_{\varphi}$ is an ${\cal N}=2$ theory with cross CS term $A_{L}dA_{R}$ at level $1$. From \cite{BDP}, we know that the application of $S$ transformation on theory ${\cal T}$ on $S^{1}\times_q D$ changes its partition function $Z_{S^{1}\times_q D}^{{\cal T}}(x)$ to
\begin{equation}
{\cal I}_{S^{1}\times_q D}^{{\cal T}}(x)\stackrel{S}{\longrightarrow}\int\frac{dz}{z}\frac{\theta(x;q)\theta(z;q)}{\theta(xz;q)}{\cal I}_{S^{1}\times_q D}^{{\cal T}}(s)
\label{S-transf-Z}
\end{equation}
We identify the kernel of integration with the index of the 2d $S$ transformation wall.
\begin{equation}\label{S-transf-W}
{\cal I}_{{\cal B}_{\varphi=S}^{(2d)}}=\frac{\theta(x;q)\theta(z;q)}{\theta(xz;q)}.
\end{equation}
From the index we can read off the field content of the theory. It has $U(1)_L\times U(1)_R$ flavor symmetry where $U(1)_R$ factor is gauged in the bulk on the right side of the wall. ${\cal B}_{\varphi=S}^{(2d)}$ consists of two Fermi multiplets and one chiral multiplet. The multiplets are charged $(1,0),(0,1)$ and $(1,1)$ under the $U(1)^{2}$ flavor symmetry. We can not extract the details of the interaction from the index as the index is insensitive to the it. It would be interesting to relate this two dimensional theory directly to the boundary term of the 3d theory with cross CS term, $\CB_{\varphi=S}$.

In the limit $q\rightarrow 1$ (\ref{S-transf-Z}) is reduced to:
\begin{equation}
 \W(x)\stackrel{S}{\longrightarrow}  \mathop{\rm ext}_z \left\{\W(z)+\log z\log x\right\},
\end{equation}
\begin{equation}
 x\longrightarrow 1/y,
\end{equation}
\begin{equation}
 y\longrightarrow x.
\end{equation}
where $\mathop{\rm ext}_z$ denotes extremization w.r.t. $z$.

Thinking of 2d transformation wall as the boundary of 3d transformation wall allows us to identify its $T^{2}$ partition function and hence, at least, its field content. This construction can be generalized by considering ${\cal T}^{(4d)}$ to be ${\cal N}=4$ SYM with gauge group $SU(N)$. On the boundary, we can consider ${\cal T}$ to be any three dimensional SCFT with $SU(N)$ flavors symmetry which is gauged by ${\cal T}^{(4d)}$ in the bulk. The three dimensional duality wall ${\cal B}_{\varphi}$ has been studied and an explicit description for both $\varphi=T,S$ in known as a three dimensional gauge theory \cite{Gaiotto:2008ak}. According to our proposal, the transformation wall ${\cal B}_{\varphi}^{(2d)}$ for theory ${\cal T}$ is simply the boundary theory of ${\cal B}_{\varphi}$.

\section{Brane models} \label{sect_branes}

In this section we consider a realization of 3d $\CN=2$ theory $\CT$ associated to a curve $\CV\subset \mathbb{C}_x^*\times \mathbb{C}_y^*$ in M-theory and type IIB string theory.

\subsection{M-theory construction}
\label{sec:Mtheorybranes}

Consider M-Theory on $\mathbb{R}^3_{016}\times \mathbb{C}_{23}\times\mathbb{C}_{45}\times\mathbb{C}_{78}^*\times\mathbb{C}_{9,10}^*$ where lower indices label corresponding real directions. In particular, the theory is compactified on the circle $S^1_y$ along the 10th direction and on $S^1_x$ along the 8th direction. Let us denote the variables parametrizing $\mathbb{C}^*_{78}$ and $\mathbb{C}_{9,10}^*$ by $x$ and $y$ respectively, so that $\mathbb{C}_{78}^*\equiv \mathbb{C}_x^*$, $\mathbb{C}_{9,10}^*\equiv \mathbb{C}_y^*$ and $(\log |x|,\log|y|)$ are coordinates on $\mathbb{R}^2_{79}$. An M2-brane on $\mathbb{R}^3_{016}$ has moduli space $\mathbb{C}^2\times(\mathbb{C}^*)^2$ as in (\ref{4d_moduli}). An M5-brane on $\mathbb{R}^2_{01}\times \mathbb{C}_{45}\times \Cs$ where $\Cs$ is considered to be embedded in $\mathbb{C}_{x}^*\times\mathbb{C}_{y}^*$ engineers a 5d theory compactified on a circle with the Seiberg-Witten curve $\Cs$ \cite{Mtheory4d,Mtheory5d}.

\begin{figure}
\centering
\includegraphics{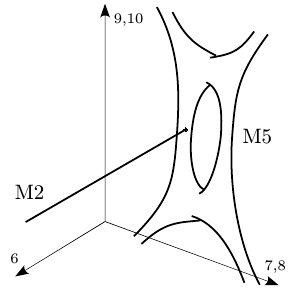}
\caption{An M2-brane ending at an M5-brane wrapping $\Cs\subset \mathbb{C}_{78}^*\times\mathbb{C}_{9,10}^*$ .}
\label{figure_M2-M5}
\end{figure}

Now let us consider a combined system: an M5-brane defined as earlier and a semi-infinite M2-brane on $\mathbb{R}^2_{01}\times \mathbb{R}_6^{+}$ ending on the M5-brane at $p\in \CV$ (see Fig. \ref{figure_M2-M5}).

This construction engineers an effective 3d theory with a 2d boundary on $\mathbb{R}^2_{0,1}$. The boundary theory depends on the position of the M2-brane in $\mathbb{C}_{45}\times \Cs$ --- the moduli space of the effective 3d theory, the same as in (\ref{4d_boundary_moduli}).

\begin{figure}
\centering
\begin{tabular}{m{0.3\textwidth}m{0.05\textwidth}m{0.3\textwidth}}
\includegraphics{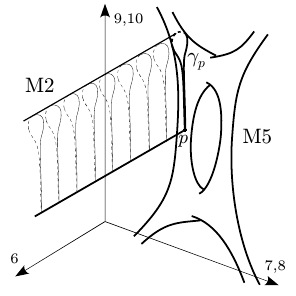} &
\Huge $\leadsto$ &
\includegraphics{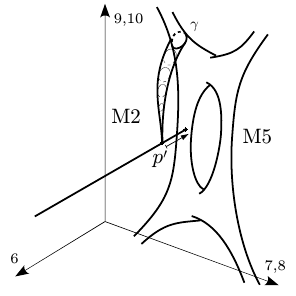}
\end{tabular}
\caption{Geometric realization of the defect associated to $\gamma_p\in  \pi_1(\Cs,p)$ in the M-theory picture.}
\label{figure_M2-M5-defect}
\end{figure}

Consider a 1d defect $L_\gamma\subset \mathbb{R}^2_{0,1}$ introduced in section \ref{sect_defects} localized in the spatial direction 1 and infinitely spread in the time direction 0. The defect can be engineered by the following geometric configuration: the position of the M2-brane in $\Cs$ changes along the contour $\gamma_p\subset\Cs\subset \mathbb{C}_{x}^*\times\mathbb{C}_{y}^*$ as we go along the spatial direction. In the limit when the width of the defect tends to zero we get an M2-brane on $\mathbb{R}_0\times \mathbb{R}_6^+\times \gamma_p$ attached to the original M2-brane on $\mathbb{R}^2_{01}\times \mathbb{R}_6^{+}$ along $L_\gamma\times \mathbb{R}_6^+$ (see Fig. \ref{figure_M2-M5-defect}).

The energy of such geometric ``excitation'' is, of course, infinite. However, one can make it finite by deforming the M2-brane on $\mathbb{R}_0\times \mathbb{R}_6^+\times \gamma_p$ to an M2-brane on $\mathbb{R}_0\times I\times \gamma$ where $I$ is a path in $\mathbb{R}_6\times \mathbb{R}^2_{79}$ connecting the M5-brane with a spatial point $p'$ on the M2-brane infinitesimally close to the boundary (see Fig. \ref{figure_M2-M5-defect}). After such deformation the position of the original M2-brane in $\mathbb{C}_{78}^*\times\mathbb{C}_{9,10}^*$ cannot be constant anymore in the vicinity of $p'$. As we go around $p'$ the M2-brane rotates by $2\pi n_\gamma$ in $\mathbb{C}^*_{9,10}\equiv \mathbb{C}^*_y$ and by $2\pi m_\gamma$ in $\mathbb{C}^*_{78}\equiv \mathbb{C}^*_x$.

\begin{figure}
\centering
\includegraphics{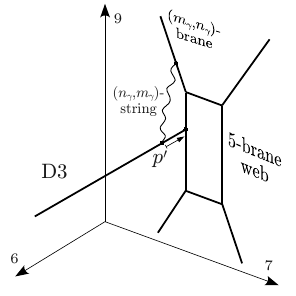}
\caption{Realization of the defect associated to $\gamma\in E\subset H_1(\Cs)$ in the type IIB string theory picture.}
\label{figure_D3-5branes}
\end{figure}

\subsection{\label{IIBstring}Type IIB string theory construction}

Let us perform compactification of M-theory on $S^1_y$ to obtain type IIA string theory and then T-duality along $S^1_x$ to obtain type IIB string theory. Type IIB string theory is realized on $\mathbb{R}^3_{016}\times \mathbb{C}_{23}\times\mathbb{C}_{45}\times \mathbb{R}_7\times\tilde{S}^1_x\times \mathbb{R}_9$. The M5-brane is translated into a web of 5-branes \cite{Mtheory4d,Mtheory5d,pqwebs} on $\mathbb{R}^2_{01}\times \mathbb{C}_{45}\times \tilde{S}^1_x\times\Cs_\trop$ where $\Cs_\trop$ is a tropical version of the curve $\Cs$ realized as a graph (representing the web) in $\mathbb{R}^2_{79}$ (see Fig. \ref{figure_D3-5branes}). The semi-infinite M2-brane ending on the M5-brane is translated into a D3-brane on $\mathbb{R}^3_{0,1}\times\mathbb{R}^+_6\times \tilde{S}_x^1$. The theory on the D3-brane is 4d $\mathcal{N}=4$ $U(1)$ SYM compactified on the circle $\tilde{S}^1_x \equiv S^1_t$ with a 3d boundary on $\mathbb{R}^2_{01}\times \tilde{S}^1_x$, as in section \ref{sect_defects}. The boundary condition given by a 5-brane web on the boundary can be understood as a generalization of the Chern-Simons boundary condition realized by a single five-brane \cite{3dpqbrane,GW2,GW3}.

Suppose we can deform the contour $\gamma$ so that in its vicinity the curve $\Cs$ locally looks like a tube $\mathbb{C}^*_z$ holomorphically embedded in $\mathbb{C}^*_x \times \mathbb{C}_y^*\equiv \mathbb{C}_{78}^*\times\mathbb{C}_{9,10}^*$ by $x\propto z^{m_\gamma}$, $y\propto z^{n_\gamma}$ (as shown in Fig. \ref{figure_M2-M5-defect}). Then in this vicinity the M5-brane translates into a $(m_\gamma,n_\gamma)$-brane --- the bound state of $m_\gamma$ NS5-branes and $n_\gamma$ D5-branes --- represented by a line with the slope $n_\gamma/m_\gamma$ in the plane $\mathbb{R}^2_{79}$. The cycles satisfying this assumption lie in the subgroup $E\subset H_1(\Cs)$. The M2-brane generating the defect translates into a $(n_\gamma,m_\gamma)$ string --- the bound state of $n_\gamma$ F1-strings and $m_\gamma$ D1-strings --- connecting the D3-brane with the $(m_\gamma,n_\gamma)$ five-brane corresponding to the cycle $\gamma$. This five-brane is supported on $\mathbb{R}^2_{01}\times \mathbb{C}_{45}\times \tilde{S}^1_x\times e$ where $e$ is the edge of $\CV_\trop$ the cycle $\gamma$ is associated to.

Let us note that the defects realized by pure F1-strings should correspond to the excitations of the chiral matter fields of the 3d boundary theory engineered by this branes configuration. The nontrivial D1-F1 bound states and D3-branes considered below are non-perturbative effects.

\begin{figure}
\centering
\begin{tabular}{m{0.3\textwidth}m{0.05\textwidth}m{0.3\textwidth}}
\includegraphics{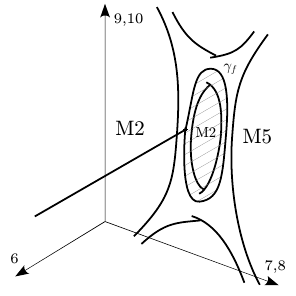} &
\Huge $\leadsto$ &
\includegraphics{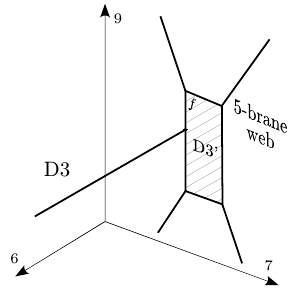}
\end{tabular}
\caption{M-theory and type IIB realization of the defect $L_{\gamma_f}$ for $\gamma_f\in F\subset H_1(\Cs_\trop)$. The cycle $\gamma_f$ corresponds to the face $f$ of the brane web.}
\label{figure_D3-5branes-D3}
\end{figure}

The type IIB realization of the defect corresponding to a cycle $\gamma_f\in F\subset H_1(\Cs)$ associated to a face $f$ of $\CV_\trop$ is shown in Fig. \ref{figure_D3-5branes-D3}. The M2-brane ending on the cycle $\gamma_f$ is translated into a D3'-brane on $\mathbb{R}_0\times\tilde{S}^1_x\times f$ where $f\subset \mathbb{R}^2_{79}$. From the point of view of the 3d boundary theory on $\mathbb{R}^2_{01}\times\tilde{S}^1_x$ this defect is two-dimensional. This provides us with another indication that the defects associated to cycles in $F\subset H_1(\Cs_\trop)$ are lifted to 2d objects in 3d. Let us note that the energy of the D3'-brane and the tropical limit of $\int_{\gamma_{f}}\log y d \log x$ are both proportional to the area of the face $f$.

To determine the effective 2d theory realized by the D3-brane stretched over the face $f$ one can use the technique of \cite{HW}. Type IIB superstring theory without branes has 32 independent supersymmetries of the form $\epsilon^L Q_L+\epsilon^R Q_R$ where $Q_L$ and $Q_R$ are supercharges generated by left- and right-moving worldsheet degrees of freedom and $\epsilon_L$ and $\epsilon_R$ are 10-dimensional spinors satisfying chirality conditions:
\begin{equation}
\begin{array}{c}
\epsilon_L=\Gamma_0\Gamma_1\Gamma_2\Gamma_3\Gamma_4\Gamma_5\Gamma_6\Gamma_7\Gamma_8\Gamma_9\epsilon_L, \\
\epsilon_R=\Gamma_0\Gamma_1\Gamma_2\Gamma_3\Gamma_4\Gamma_5\Gamma_6\Gamma_7\Gamma_8\Gamma_9\epsilon_R.
\end{array}
\label{epsilons-10dchirality}
\end{equation}
Introducing a D3'-brane along the directions $0879$ reduces the number of supersymmetries to 16 by imposing the following condition:
\begin{equation}
\text{D3':}\qquad\epsilon_L=\Gamma_0\Gamma_7\Gamma_8\Gamma_9\epsilon_R.
\label{epsilons-D3p}
\end{equation}
The degrees of freedom of the D3'-brane in the absence of any other branes are given by $\CN=4$ 4d vector multiplet containing a vector field $A_\mu$ and 6 real bosonic fields $\phi_1,\phi_2,\phi_3,\phi_4,\phi_5,\phi_6$ describing the position of the brane. Adding NS5- and D5-branes (and/or their bound states) bounding the D3-brane imposes additional conditions which reduce the number of supersymmetries to 4:
\begin{equation}
\begin{array}{c}
\text{NS5:}\qquad \epsilon_L=\Gamma_0\Gamma_1\Gamma_4\Gamma_5\Gamma_7\Gamma_8\epsilon_L,\;
\epsilon_R=\Gamma_0\Gamma_1\Gamma_4\Gamma_5\Gamma_7\Gamma_8\epsilon_R,
\\
\text{D5:}\qquad
\epsilon_L=\Gamma_0\Gamma_1\Gamma_4\Gamma_5\Gamma_8\Gamma_9\epsilon_R.
\end{array}
\label{epsilons-NS5-D5}
\end{equation}

One can check that all solutions of (\ref{epsilons-10dchirality}),(\ref{epsilons-D3p}) and (\ref{epsilons-NS5-D5}) satisfy the following chirality conditions in the 2d space-time spanned along the directions $08$:
\begin{equation}
\epsilon_L=\Gamma_0\Gamma_8\epsilon_L,\;
\epsilon_R=\Gamma_0\Gamma_8\epsilon_R.
\end{equation}
It follows that the effective 2d theory engineered by the D3-brane bounded by 5-branes has $(0,4)$ supersymmetry. The remaining massless degrees of freedom\footnote{While the massless degrees of freedom do not depend on the shape of the face, one should expect that the spectrum of massive Kaluza-Klein-like modes does depend on it.} are given by a $(0,4)$ chiral multiplet containing $A_9,\phi_1,\phi_4,\phi_5$ as bosonic components. Finally, adding a D3-brane along the directions $0168$ reduces the supersymmetry to $(0,2)$ by the following additional condition:
\begin{equation}
\text{D3:}\qquad\epsilon_L=\Gamma_0\Gamma_1\Gamma_6\Gamma_8\epsilon_R.
\label{epsilons-D3}
\end{equation}
The condition that D3'-brane intersects the boundary of the D3-brane along the 2-dimensional subspace $08$ fixes the position of the D3'-brane in the directions $45$. This reduces the $(0,4)$ multiplet to a $(0,2)$ chiral multiplet containing $A_9,\phi_1$ as bosonic fields.

Let us note that a related brane model of $(0,2)$ theories was considered in \cite{BraneBox}. One can perform T-duality along the direction 4, then lift to M-theory by introducing an extra direction 6' and then compactify on a circle along the 6th direction to transform the branes in the following way:
\begin{equation}
\begin{array}{lcl}
\text{NS5 along 014578 }&\rightarrow &\text{ NS5 along 014578}\\
\text{D5 along 014589 }&\rightarrow &\text{ NS5' along 0156'89}\\
\text{D3' along 0789 }&\rightarrow &\text{ NS5'' along 046'789}\\
\text{D3 along 0168 }&\rightarrow &\text{ D4 along 0146'8}\\
\end{array}
\label{boxdual}
\end{equation}
We end up with the same brane setup as was considered in \cite{BraneBox}.

\subsection{Relation to the standard brane model} \label{sect_branes_usual}

In this section we describe how this brane construction is related to the brane model for $\mathcal{N}=2$ 3d (or $\mathcal{N}=(2,2)$ 2d) theories considered previously in the literature (see e.g. \cite{HW,DT,3dpqbrane,HH,BHOY,DGH}).

\begin{figure}
\centering
\begin{tabular}{m{0.27\textwidth}m{0.03\textwidth}m{0.65\textwidth}}
\includegraphics{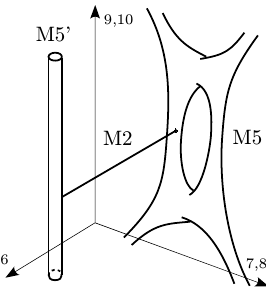} &
\Huge $\leadsto$ &
\includegraphics{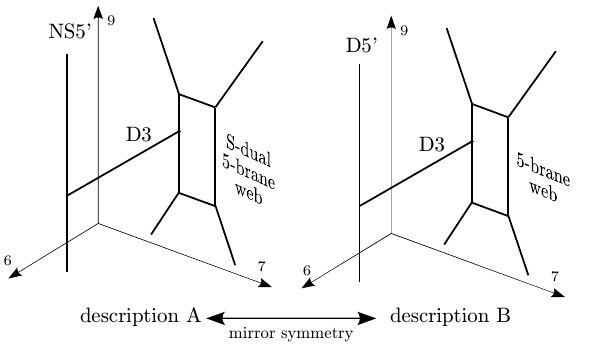}
\end{tabular}
\caption{An M2-brane stretched between an M5-brane wrapping $\Cs\subset \mathbb{C}_x^*\times\mathbb{C}_y^*$ and an M5'-brane wrapping $\mathbb{C}_y^*\subset \mathbb{C}_x^*\times\mathbb{C}_y^*$. In the type IIB string theory picture it is translated into D3 brane stretched between a five-brane web and an NS5'-brane or between the S-dual five-brane web and a D5'-brane depending on the choice of the compactification circle.}
\label{figure_M2-M5-M5}
\end{figure}

Let us start again with M-theory on $\mathbb{R}^3_{016}\times \mathbb{C}_{23}\times\mathbb{C}_{45}\times\mathbb{C}_x^*\times\mathbb{C}_y^*$ and an M5-brane wrapping $\mathbb{R}^2_{01}\times \mathbb{C}_{45}\times \Cs$. Now, instead of a semi-infinite M2-brane let us consider an M2-brane with a finite extension in the direction 6 stretched between the M5-brane and an auxiliary M5'-brane on $\mathbb{R}^2_{01}\times \mathbb{C}_{23}\times \mathbb{C}_y^*$ where $\mathbb{C}_y^*$ is embedded in $\mathbb{C}_x^*\times\mathbb{C}_y^*$ by fixing a particular value of $x$ (see Fig. \ref{figure_M2-M5-M5}). Since the M2-brane has a finite extension in the direction 6 the effective theory is now purely two-dimensional (instead of three-dimensional with a boundary) with the spacetime $\mathbb{R}^2_{0,1}$. Fixing the value of $x$ of the M5'-brane corresponds to a particular choice of the SUSY parameter $x$ of this theory. After introducing an M5'-brane along $\mathbb{C}_x^*$ one direction in $\mathbb{C}_x^*\times \mathbb{C}_y^*$ becomes distinguished which corresponds to the choice of polarization considered in \cite{DGSdual,DGG1}.

If $S_y^1$ plays the role of the M-theory circle, as in the previous section, the M5'-brane becomes a D5-brane (which we will denote as D5') on $\mathbb{R}^3_{016}\times \tilde{S}^1_x$ in the type IIB picture (see Fig. \ref{figure_M2-M5-M5}, description B). Now let us consider instead $S_x^1$ as the M-theory circle. The M2-brane is then translated into a D3-brane on $\mathbb{R}^3_{016}\times \tilde{S}^1_y$. The M5-brane becomes the same 5-brane web as in the previous section but with D5 and NS5 branes interchanged. The 5-branes of the web are supported on $\mathbb{R}^2_{01}\times \mathbb{C}_{45}\times \tilde{S}^1_y\times\Cs_\trop$. The M5'-brane becomes an NS5'-brane on $\mathbb{R}^2_{01}\times \mathbb{C}_{23}\times \tilde{S}^1_y\times \mathbb{R}_7$ (see Fig. \ref{figure_M2-M5-M5}, description A).

The position of the NS5'/D5'-brane in the direction 7 equals $\log|x|$. The brane construction in the description A can be understood as a generalization of the standard brane construction \cite{HW,3dpqbrane,BHOY,DT} of a $\mathcal{N}=2$ $U(1)$ gauge theory where $\log|x|$ can be interpreted as FI parameter and the vertical position (the value of $\log|y|$) of the D3-brane as the v.e.v. of the scalar field from the vector multiplet. D5-branes in the web correspond to chiral multiplets charged w.r.t. this $U(1)$. This assumption means that the effective twisted superpotential has the form
\begin{equation}
 \W(z;x)=f(z)+\log z \log x
\end{equation}
where the variable $z$ corresponds to the twisted chiral multiplet constructed from the $U(1)_z$ vector multiplet. Then indeed on-shell we have
\begin{equation}
 y:= e^{x\frac{\d \W}{\d x}}= z.
\end{equation}

Let us consider the following simple illustrative example: SQED with $N_f'$ chiral multiplets with charge $+1$ and  $N_f$ chiral multiplets with charge $-1$. Let the \textit{effective} CS level (see e.g. \cite{aspects3d}) be $k$. Then
\begin{equation}
 \W(z;x)=\sum_{i=1}^{N_f}\dilog(ze^{-\tilde{m}_i})-\sum_{i=1}^{N_f'}\dilog(ze^{-m_i})+\frac{k}{2}(\log z)^2+\log z \log x.
 \label{W-SQED-Nf}
\end{equation}
Where $m_i$ and $-\tilde{m}_i$ are complex parameters --- combinations of real mass parameters and R-charges.
The curve $\CV$ is given by
\begin{equation}
 \prod_{i=1}^{N_f}(1-ye^{-\tilde{m}_i})=xy^k\prod_{i=1}^{N_f'}(1-ye^{-{m}_i}).
\end{equation}
\begin{figure}
\centering
\includegraphics{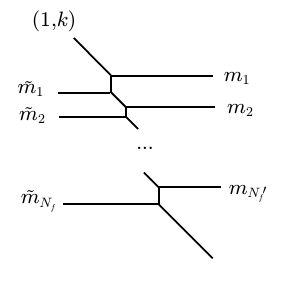}
\caption{The brane web for SQED with $N_f^{\prime}$ chirals with charge $+1$ and $N_f$ chirals with charge $-1$. The chiral fields correspond to the fundamental strings stretched between the horizontal semi-infinite D5-branes and the D3-brane probing the web.}
\label{figure_SQED-Nf-branes}
\end{figure}

The brane web is shown in Fig. \ref{figure_SQED-Nf-branes}. If it happens that $m_i=\tilde{m}_j$ for some $i$ and $j$ then the curve becomes reducible with a separate component given by
\begin{equation}
 y=e^{m_i}.
\end{equation}
This equation describes an infinite horizontal D5 brane producing a hypermultiplet. Moving this D5-brane away from the web along $\mathbb{C}_{23}$ corresponds to introducing a complex mass (see e.g. \cite{BHOY}).

The theory produced by the brane model in the description B does not have a gauge field. The value of $\log|x|$ can be interpreted as a real mass parameter shift for chiral fields corresponding to D5-branes (which have vertical orientation) in the web. The simple class of such theories can be obtained by $S$-transformation (which corresponds to rotation of the 5-brane web by $\pi/2$ combined with S-duality) of SQED:

\begin{equation}
 \W(x)=\sum_{i=1}^{N_f}\dilog(xe^{-\tilde{m}_i})-\sum_{i=1}^{N_f'}\dilog(xe^{-m_i})+\frac{k}{2}(\log x)^2.
 \label{W-ndSQED}
\end{equation}
This theory contains $N_f'$ chiral multiplets with charge $+1$ w.r.t. the global symmetry $U(1)_x$ and $N_f$ chiral multiplets with charge $-1$.
The curve in this case is the following:
\begin{equation}
 \prod_{i=1}^{N_f}(1-xe^{-\tilde{m}_i})y=x^k\prod_{i=1}^{N_f'}(1-xe^{-{m}_i}).
\end{equation}

The relation between the curve $\CV$ and the brane-web given by $\CV_\text{trop}$ can be interpreted in terms of topological strings \cite{BranesToric}. Let us consider A-model topological strings on local Calabi-Yau three-fold whose toric diagram is given by $\CV_\trop$ with complexified Kähler parameters $\log t_i$ --- length parameters of $\CV_\trop$. In the mirror B-model the Calabi-Yau is defined by the equation $A(x,y)=uv$. Then, in general, if the D3-brane is probing any non-vertical 5-brane with the slope $k$ there is a realization of $\W$ given by the topological string disk amplitude \cite{OV,AVmirror}:
\begin{equation}
 \W(x)=\frac{k}{2}(\log x)^2+c\log x+\sum_{\alpha,\beta}N_{\alpha,\beta}\dilog(x^{-\alpha}\prod_i t_i^{-\beta_i}),
 \qquad |x^{-\alpha}\prod_i t_i^{-\beta_i}|<1
 \label{top-non-vert}
\end{equation}
The parameter $x$ plays the role of the modulus of the Lagrangian brane defining the boundary condition for the disk. Where $N_{\alpha,\beta}$ are integers. This realization of $\W$ corresponds to the type IIB brane description B. The theory can be interpreted as a CS theory for a non-dynamical vector multiplet associated to the global symmetry $U(1)_x$ with some number (possibly infinite) of chiral fields with masses given by linear combinations of $\log x$ as well as $\log t_i$.  The theory (\ref{W-ndSQED}) is a particular example of such description (for the region where $x$ is sufficiently small).

 Analogously, for any non-horizontal brane there is a description with a dynamical $U(1)_z$ gauge field corresponding to the type IIB brane description A:
\begin{equation}
 \W(z;x)=\frac{k'}{2}(\log z)^2+c'\log z+\sum_{\alpha,\beta}N'_{\alpha,\beta}\dilog(z^{-\alpha}\prod_i t_i^{-\beta_i})+\log z\log x,
 \qquad |z^{-\alpha}\prod_i t_i^{-\beta_i}|<1.
 \label{top-non-hor}
\end{equation}
Here $\log x$ plays the role of the FI parameter. For example, such description for the lower semi-infinite leg in Fig.~\ref{figure_SQED-Nf-branes} gives (\ref{W-SQED-Nf}). 

If we restore dependence on the compactification radius (see e.g. \cite{NS}) the contribution from the chiral field has the following form:
\begin{equation}
 \frac{1}{R}\dilog (e^{-RM}).
\end{equation}
If $M>0$ this contribution is exponentially suppressed in the decompactification limit $R\rightarrow\infty$. Therefore in the 3d limit only quadratic terms survive in the representations of type (\ref{top-non-vert}) or (\ref{top-non-hor}). This limit can be understood as the tropical limit because, as usual, one can obtain the tropical curve $\CV_\trop$ from the ordinary one $\CV$ given by the equation $A(x,y)=0$ by variabeles $(x,y)\in \mathbb{C}^*\times\mathbb{C}^*$ and all parameters $t_i\in \mathbb{C}^*$ in the form $x=e^{RX}$, $y=e^{RY}$, $t_i=e^{RT_i}$ and taking the limit $R\rightarrow +\infty$. In the tropical limit $\log y$ is a locally linear function of $\log x$ which means that $\W(x)$ is locally quadratic, that is, corresponds to CS theory in 3d. The similar situation happens when $R$ is fixed but $M\rightarrow\infty$ for all chiral fields. This happens at the ends of non-compact edges of $\Cs_\trop$. This means that at singular points of the curve $\Cs$ the theory admits purely CS description without chiral fields.

In the 3d limit the situation when the D3-brane is stretched between NS5'- and NS5-branes which have the same horizontal position (i.e. $\log x$) corresponds to unbroken $U(1)_z$ gauge symmetry.

Theories constructed by brane models A and B (see Fig.~\ref{figure_D3-5branes-D3}) are related by mirror symmetry, although in general one (or both) of the theories might not have a good Lagrangian (with finite number of chiral multiplets). While mirror symmetry can be understood as S-duality of the brane model, $S$-transformation considered in section \ref{sect_transf_walls} can be realized by S-duality \textit{and} rotation of the brane web by $\pi/2$ (keeping other branes intact) which in general produces an essentially different theory.

\section{Prominent examples} \label{sect_examples}

Before describing particular examples let us first consider a general situation when the 3d $\mathcal{N}=2$ theory $\CT$ has Lagrangian description with a finite number of chiral fields and  abelian symmetries $\{U(1)_{\xi}\,|\,\xi\in \{x\}\cup\{t_i\}\cup\{z_j\}\}$ where $U(1)_{x}$ and $U(1)_{t_i}$ are global symmetries and $U(1)_{z_j}$ are gauge symmetries. The scalar component of the twisted chiral field $\Sigma_\xi$ corresponding to the $U(1)_\xi$ vector field is $\log\xi$. Suppose the theory has chiral fields $Q_\ell$ with the charges $n_{\xi,\ell}$ w.r.t. $U(1)_{\xi}$ and complexified mass parameters $n_{x,\ell}\log x+\sum_i n_{t_i,\ell}\log t_i$. Then the effective twisted superpotential has the following form:

\begin{equation}
 \W (z_i;x,t_i)=\sum_{\ell}\dilog\left(\prod_\xi \xi^{-n_{\xi,\ell}}\right)+\frac{1}{2}\sum_{\xi,\tilde{\xi}}k_{\xi,\tilde{\xi}}\log \xi \log\tilde\xi
 \label{W-general-lagrangian}
\end{equation}
The \textit{polynomial} equations
\begin{equation}
\begin{array}{c}
 y=e^{x\frac{\d \W}{\d x}} \\
 1=e^{z_i\frac{\d \W}{\d z_i}}
\end{array}
 \label{curveeqs}
\end{equation}
define a curve $\CV'\subset (\mathbb{C}^*)^{s+2}$ where $s$ is the number of $U(1)_{z_i}$ gauge symmetries. Using resultants one can always eliminate all $z_i$ and reduce (\ref{curveeqs}) to an equation for two variables defining a plane curve $\CV$:
\begin{equation}
 A(x,y)=0.
\end{equation}
Geometrically this corresponds to the projection $\CV'\to \CV\subset (\mathbb{C}^*)^2$. We will make the following assumption (which holds in generic situation): this projection is a map of degree 1, i.e. it is almost everywhere injective and provides a birational equivalence between $\CV'$ and $\CV$. Algebraically it corresponds to the fact that $A(x,y)\neq (B(x,y))^k$ for some $k>1$ and a polynomial $B(x,y)$. Then  $z_i$ can be considered as (almost everywhere) single-valued functions on the curve $\CV$. When the point $p\in \CV$ goes along a closed cycle, $z_i(p)$ also makes a closed cycle in $\mathbb{C}^*_{z_i}$. The monodromies of $\W$ then can be computed using the following basics rules:
\begin{equation}
\begin{array}{c}
  \mon\limits_{\gamma_{\zeta=0}} \log\zeta=2\pi i \\
 \mon\limits_{\gamma_{\zeta=1}}\dilog(\zeta)=2\pi i\log\zeta
\end{array}
\label{mon_rules}
\end{equation}
where $\gamma_{\zeta=\zeta_0}$ is a cycle with the winding number $+1$ around $\zeta_0$ in the complex plane parametrized by $\zeta$. The monodromy of the effective twisted superpotential has therefore the following simple form:
\begin{equation}
\mon_\gamma\W=2\pi^2q_0+2\pi i\sum_\xi q_\xi\log\xi,\qquad q_0,\,q_\xi\in\mathbb{Z}.
\end{equation}
Using that the dependence on the starting/ending point of the contour should be of the form (\ref{Weff_monodromy_res}) it is easy to see that the monodromy of $\W$ for \textit{any} cycle has the form\footnote{We also assume that there is no trivial relation between $z_i$'s of the form $\prod_i z_i^{\ell_i}\propto x$ and therefore all $q_{z_i}$ must be zero.} (\ref{periods_moduli}). In the remainder of the section we consider several illustrative examples, for more examples see appendix \ref{example-table}.

\subsection{Supersymmetric Chern-Simons theory}

Consider $\mathcal{N}=2$ Chern-Simons theory with level $k$ for the background $U(1)_x$ vector multiplet:
\begin{equation}
 \W (x)=\frac{k}{2}(\log x)^2.
\end{equation}
The curve $\CV$ is given by:
\begin{equation}
 y=x^k.
\end{equation}
The web in the type IIB brane model contains a single $(1,k)$ brane. There is only one non-trivial cycle on the curve, $\gamma$:
\begin{equation}
\begin{array}{rlcrl}
\mon_{\gamma}\W =& 2\pi i k\log x-2\pi^2 k,
&\qquad & c_*\gamma= & (k,1).\\
\end{array}
\end{equation}
The $S$-transformed theory is given by
\begin{equation}
 \W (z;x)=\frac{k}{2}(\log z)^2+\log z\log x,
\end{equation}
which after extremization w.r.t. $z$ is reduced to
\begin{equation}
 \W (z;x)=-\frac{1}{2k}(\log x)^2.
\end{equation}
Thus $S$-transformation of CS-theory effectively inverts CS level: $k\rightarrow -1/k$ (cf. \cite{3dpqbrane}).

\subsection{Theory $\CT_\Delta$}
\begin{figure}
\centering
\includegraphics{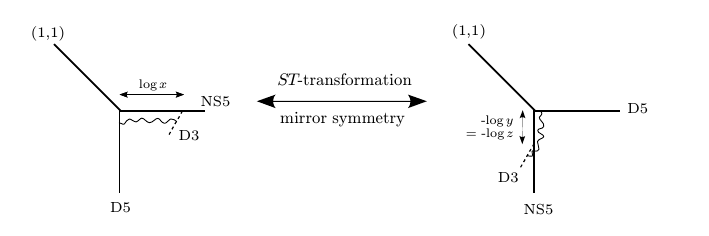}
\caption{Different descriptions of the theory $\CT_\Delta$. The fundamental string stretched between D3 and D5 corresponds to the chiral field neutral in the first case and charged in the second case. For the first case $\log x$ plays the role of a mass parameter, in the second case $\log x$ is the FI parameter.}
\label{figure_T_Delta}
\end{figure}

Let us consider one chiral multiplet with charge $+1$ w.r.t. $U(1)_x$ and complexified mass parameter $\log x$ and zero \textit{effective} CS level (the bare CS level is $-1/2$):
\begin{equation}
\W(x)=\dilog(x^{-1}).
\label{T_Delta_B}
\end{equation}
The curve is the following:
\begin{equation}
 y=1-x^{-1}.
\end{equation}
In \cite{DGSdual,DGG1,DGGindex} this theory was considered as the one corresponding to a tetrahedron, so that $\W(x)$ equals to its hyperbolic volume. The theory is well defined in the region $|x|>1$ (in the limit when the mass is very large $\Re\log x\rightarrow\infty$ the chiral field decouples) corresponding to the horizontal semi-infinite brane of the brane model shown on the left of Fig. \ref{figure_T_Delta}. Following the general discussion in section \ref{sect_branes_usual} there is a mirror description of the theory shown on the right side of Fig. \ref{figure_T_Delta}. The mirror theory is $U(1)$ gauge theory with FI-parameter $\log x$ and one massless charged chiral multiplet. It has the following effective twisted superpotential (cf. \cite{DGG1}):
\begin{equation}
 \W(x;z)=-\dilog(z)+\log z\log x=\dilog(z^{-1})+\frac{1}{2}(\log z)^2+\log z\log x
 \label{T_Delta_A}
\end{equation}
where for the sake of simplicity we suppressed linear and constant terms. Comparing (\ref{T_Delta_A}) and (\ref{T_Delta_B}) one can see that the mirror descriptions are related by $ST$-transformation \cite{DGG1,DGGindex}. This corresponds to the fact that the brane web (see Fig. \ref{figure_T_Delta}) of this particular theory is \textit{invariant} under $ST\in SL(2,\mathbb{Z})$ transform in the plane $\mathbb{R}^2_{79}$ parametrized by $(\log|x|,\log|y|)$.

On the quantum level this relation corresponds to the following identity \cite{BDP}:
\begin{equation}
 (q^{-1})_\infty^{-1}\CI^{\CT_\Delta}_{S^1\times D}(x)=\int \frac{dz}{z}\frac{\theta(-q^{1/2}x;q)}{\theta(-q^{1/2}xz;q)}\CI^{\CT_\Delta}_{S^1\times D}(z)
\label{q_dilog_identities_2}
\end{equation}
where
\begin{equation}
 \CI^{\CT_\Delta}_{S^1\times D}(x)=(qx^{-1};q)_\infty,\qquad (\xi;q)_\infty:= \prod_{i=0}^\infty(1-\xi q^i)
\end{equation}
is the ``half-index" for the theory $\CT_\Delta$. The right hand side of (\ref{q_dilog_identities_2}) is given by the composition of $T$ and $S$ transformations considered in section \ref{sect_transf_wall_3d} (up to rescalings).

The curve of the theory has two independent cycles associated to the horizontal and vertical semi-infinite branes:
\begin{equation}
\begin{array}{rlcrl}
\mon_{\gamma_1}\W =& 0,
&\qquad & c_*\gamma_1= & (0,1), \\
\mon_{\gamma_2}\W =& -2\pi i\log x ,
&\qquad & c_*\gamma_2= & (1,0). \\
\end{array}
\end{equation}
Let us note that the electron of $U(1)_z$ is the monopole of $U(1)_x$ and vice versa. It is easy to see that the masses of particles corresponding to $\gamma_{1,2}$ are indeed given by $|\mon_{\gamma_{1,2}}\W/2\pi i|$.

\subsection{$\CN=2$ SQED and XYZ theory}

Let us consider the theory introduced in section \ref{sect_branes_usual} with $N_f=N_f'=1$, namely SQED with two chiral fields with charges $1$ and $-1$ and mass $m$:
\begin{equation}
 \W(z;x,m)=\dilog(ze^{-m})-\dilog(ze^{m})-m\log z+\log x \log z.
  \label{W-SQED-1}
\end{equation}
The curve is given by:
\begin{equation}
 x(1-ye^m)=(e^m-y).
\end{equation}
Or, equivalently:
\begin{equation}
 y(1-xe^{m})=(e^m-x).
\end{equation}
From this representation one can see that the theory is equivalent to one of the form (\ref{W-ndSQED}) and has an alternative description without a dynamical vector field:
\begin{equation}
 \W(x,m)=\dilog(xe^{m})-\dilog(xe^{-m})+m \log x.
 \label{W-SQED-2}
\end{equation}

\begin{figure}
\centering
\begin{tabular}{m{0.3\textwidth}m{0.05\textwidth}m{0.3\textwidth}}
\includegraphics{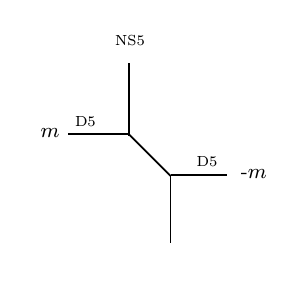} &
\Huge $\cong$ &
\includegraphics{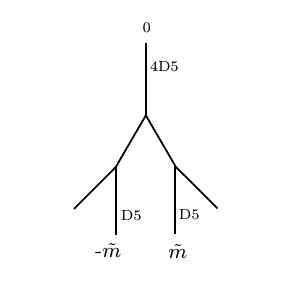}
\end{tabular}
\caption{The left picture is provides a natural brane description of SQED (with two charged chiral fields corresponding to two horizontal semi-infinite D5-branes) if we treat $\log x$ as the horizontal coordinate and $m$ as a parameter of the brane web. The right picture provides a natural brane description of XYZ model if we interpret $\log \tilde x= m$ as the horizontal coordinate. Three vertical branes correspond to three chiral fields charged w.r.t. $U(1)_{\tilde{x}}$. The upper semi-infinite brane can be considered as a bound state of 4 D5-branes.}
\label{figure_SQED-XYZ}
\end{figure}

There are 4 independent cycles on the curve:
\begin{equation}
\begin{array}{rlcrl}
\mon_{\gamma_1}\W =& 2\pi i m,
&\qquad & c_*\gamma_1= & (0,1), \\
\mon_{\gamma_2}\W =& 2\pi i m ,
&\qquad & c_*\gamma_2= & (0,-1), \\
\mon_{\gamma_3}\W =& 2\pi i\log x+2\pi i m ,
&\qquad & c_*\gamma_3= & (1,0), \\
\mon_{\gamma_4}\W =& -2\pi i\log x+2\pi i m ,
&\qquad & c_*\gamma_4= & (-1,0), \\
\end{array}
\end{equation}
where $\gamma_{1,2}$ and $\gamma_{3,4}$ are associated to the horizontal and the vertical legs of the brane web (see Fig. \ref{figure_SQED-XYZ}), respectively.

One can also construct a curve using $\tilde{x}:= e^m$ as a distinguished $\mathbb{C}^*$ parameter (and treating $\tilde{m}=:\log x$ as a modulus of the curve):
\begin{equation}
 \tilde{y}(1-\tilde{x}e^{\tilde{m}})(1-\tilde{x}e^{-\tilde{m}})=-\tilde{x}^{-1}(1-\tilde{x}^2)^2.
\end{equation}
This indicates that the theory has description of the form (\ref{W-ndSQED}):
\begin{multline}
 \W(\tilde{x},\tilde{m})=\dilog(\tilde{x}e^{\tilde{m}})+\dilog(\tilde{x}e^{-\tilde{m}})
 -2\dilog(\tilde{x})-2\dilog(-\tilde{x})-\frac{1}{2}(\log(-\tilde{x}))^2\\
 =\dilog(\tilde{x}e^{\tilde{m}})+\dilog(\tilde{x}e^{-\tilde{m}})
 -\dilog(\tilde{x}^2)-\frac{1}{2}(\log(-\tilde{x}))^2
  \label{W-SQED-3}
\end{multline}
which is $\W$ for XYZ model \cite{DGG1}. We used the following peculiar identity:
\begin{equation}
 2\dilog(\tilde{x})+2\dilog(-\tilde{x})=\dilog(\tilde{x}^2)
\end{equation}
which can be understood as equivalence of a combination of 4 chiral fields to a single chiral field. In the brane model it corresponds to combining 4 branes into one (see Fig. \ref{figure_SQED-XYZ}). It is easy to see that (\ref{W-SQED-2}) agrees with (\ref{W-SQED-3}) since (\ref{W-SQED-2}) is defined up to a function of $m$ and (\ref{W-SQED-3}) is defined up to a function of $\tilde{m}$.

\subsection{$\mathbb{CP}^1$ model}
Consider SQED with two chiral fields with charge $+1$ and masses $\pm m$:
\begin{equation}
 \W(z;x)=\frac{1}{2}(\log z)^2+\frac{1}{2}m^2+\log(-x)\log z+\dilog(z^{-1}e^m)+\dilog(z^{-1}e^{-m})
\end{equation}
The curve $\CV$ is given by the following equation:
\begin{equation}
 -x(1-ye^{-m})(1-ye^{m})=y.
\end{equation}
In general for a given $x$ there are two possible values of $y$ corresponding to two different vacua.

\begin{figure}
\centering
\includegraphics{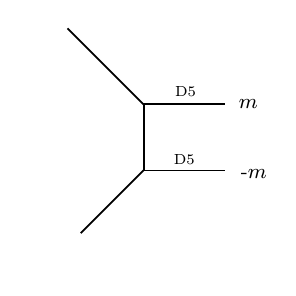}
\caption{The brane web description of SQED with two charged chiral fields corresponding to two horizontal semi-infinite D5-branes (cf. \cite{DT})}
\label{figure_CP1}
\end{figure}

The curve has 4 independent cycles associated to the external legs of the brane web. See Fig. \ref{figure_CP1}:
\begin{equation}
\begin{array}{rlcrl}
\mon_{\gamma_1}\W =& 2\pi i m,
&\qquad & c_*\gamma_1= & (0,1), \\
\mon_{\gamma_2}\W =& -2\pi i m ,
&\qquad & c_*\gamma_2= & (0,1), \\
\mon_{\gamma_3}\W =& -2\pi i\log x ,
&\qquad & c_*\gamma_3= & (-1,1), \\
\mon_{\gamma_4}\W =& 2\pi i\log x ,
&\qquad & c_*\gamma_4= & (1,1). \\
\end{array}
\end{equation}

There is no simple Lagrangian description of the mirror theory\footnote{Here, as in the rest of the paper, we understand mirror symmetry as described in the section \ref{sect_branes_usual}, that is duality between theories with and without $U(1)$ gauge symmetry corresponding to the S-duality acting on the type IIB brane construction. It differs from the mirror symmetry of $\mathbb{CP}^1$ models considered in \cite{BDP}. }. However (in accordance with the general discussion in section \ref{sect_branes_usual}) there is a description with infinite number of chiral multiplets. For the D3-brane probing upper (lower) semi-infinite horizontal 5-brane we have
\begin{equation}
 \W(x)=\pm m\log x\pm \sum_{k,j> 0}N_{k,j}\dilog(x^{-k}e^{-2jm}),\qquad N_{k,j}\in \mathbb{Z}_{\geq 0}
\end{equation}
up to a function of $m$.


\section{Duality between 3d $\CN=2$ theories via integrable models}
\label{sec:integrability}

Motivated by the relation \cite{DGH,DGSdual} to similar structures in $SL(2,\C)$ Chern-Simons theory \cite{Apol,DGLZ,Tudor}, it was recently emphasized that a convenient and useful way to describe the supersymmetric partition functions (\ref{ZZZasymp}) is by writing $q$-difference recursion relations that they satisfy. Such recursion relations can be written in the operator form \cite{DGG1,DGGindex,BDP}:
\begin{equation}
 \hat{A}(\hat x,\hat y;q) \, Z (\CT; q,x) \; = \; 0
 \label{quantumcurve}
\end{equation}
and interpreted as Ward identities for Wilson and 't Hooft lines in 4d gauge theory coupled to the 3d $\CN=2$ theory $\CT$. Specifically, at the level of partition functions (\ref{ZZZasymp}), incorporating Wilson and 't Hooft line operators is described by operators $\hat x$ and $\hat y$, respectively:
\begin{equation}
Z (\CT + \text{Wilson};q,x) \; = \; \hat x Z (\CT; q,x) \; = \; x Z (\CT; q,x) \,,
\label{Wxacts}
\end{equation}
\begin{equation}
Z (\CT + \text{'t Hooft}; q,x) \; = \; \hat y Z (\CT; q,x) \; = \; Z (\CT; q,qx) \,,
\label{Hyacts}
\end{equation}
which in the equivariant setting (a.k.a. non-trivial $\Omega$-background) generate a non-commutative algebra \cite{ramified},
\begin{equation}
 \hat y \hat x \; = \; q \hat x \hat y \,.
 \label{xycomm}
\end{equation}
Therefore, Ward identities for line operators in a coupled 3d-4d system can be represented by (polynomial) equations in $\hat x$ and $\hat y$, similar to \eqref{Vxiyi}. In fact, in the limit $q \to 1$ these equations describe the same algebraic variety $\CV$ that played a key role in this paper, and that in the simplest case of $n=1$ symmetry group $U(1)_x$ is just an algebraic curve \eqref{Acurve}. Put differently, the operator equation \eqref{quantumcurve} is a ``quantum version'' of the algebraic curve \eqref{Acurve} associated to a 3d $\CN=2$ theory $\CT$.

The discussion in this section is based on the following observation.\footnote{which already played some role in the existent literature \cite{MM2012}, albeit in a different context} The quantum algebraic curves \eqref{quantumcurve} of 3d $\CN=2$ theories have the same general form as the Baxter equation of {\it trigonometric} (sometimes also called {\it hyperbolic}) integrable systems, such as the XXZ spin chain or the trigonometric case of the Ruijsenaars model. Therefore, to every such integrable system we can associate a 3d $\CN=2$ theory $\CT_{\text{here}}$ with a distinguished flavor symmetry $U(1)_x$, such that the Ward identity for line operators in the 3d theory $\CT_{\text{here}}$ is identical to the Baxter equation of the corresponding integrable system:

\begin{equation}
\renewcommand{\arraystretch}{1.3}
\begin{tabular}{|@{\quad}c@{\quad}|@{\quad}c@{\quad}| }
\hline  {\bf Integrable System} & {\bf 3d} $\CN=2$ {\bf theory} $\CT_{\text{here}}$
\\
\hline
\hline \multirow{2}{*}{Baxter equation} &  Ward identity \\
 & for line operators \\
\hline spectral curve $A(x,y)=0$ & SUSY parameter space $\CV$ \\
\hline Quantization & $\Omega$-deformation in \\
$q = e^{\hbar}$ & 3d space-time $S^1 \times_q \R^2$ \\
\hline
\end{tabular}
\label{thedictionary}
\end{equation}

\subsection{Does S stand for ``spectral''?}

We call this theory $\CT_{\text{here}}$ to distinguish it from another 3d $\CN=2$ theory $\CT_{\text{NS}}$ associated to the same type of integrable system via Nekrasov-Shatashvili duality \cite{NS}:
$$
\boxed{\phantom{\int} \CT_{\text{NS}} \phantom{\int}}
\quad \xleftarrow[~]{\text{Nekrasov-Shatashvili}} \quad
\boxed{{\text{Integrable} \atop \text{System}}}
\quad \xrightarrow[\text{identity for line operators}]{\text{Baxter equation as Ward}} \quad
\boxed{\phantom{\int} \CT_{\text{here}} \phantom{\int}}
$$
Curiously, the two 3d $\CN=2$ theories $\CT_{\text{here}}$ and $\CT_{\text{NS}}$ associated to the same integrable system in general are {\it not} the same, as we explain in more detail below.\footnote{\label{foot:adj}For instance, one of the basic reasons is that the quantization parameter $q=e^{\hbar}$ in the present approach comes from the $\Omega$-deformation along the 3d space-time $S^1 \times_q \R^2$ of the $\CN=2$ theory $\CT_{\text{here}}$, whereas in the Nekrasov-Shatashvili duality $\hbar = m^{\text{adj}}$ is the mass of the adjoint matter multiplet, and no $\Omega$-deformation is required \cite{NS}. This will be discussed in more details below.} Therefore, a relation between the left-hand side and the right-hand side of this diagram can be viewed as a non-trivial map between 3d $\CN=2$ theories, which goes via integrable systems in the middle.

In what follows, we refer to this map between $\CT_{\text{here}}$ and $\CT_{\text{NS}}$ as a {\it 3d spectral duality} since the spectral curve $\CV$ is one of the main ingredients that provides the link. Even though this name is chosen by analogy with the spectral duality between integrable models \cite{AHH,Wilson,Ruijsenaars,BertolaEH,MTVbi,MTV2012,Morozov-duality,Sasha-duality} --- which is also based on identification of spectral curves and whose origin goes back to mirror symmetry \cite{Givental} and Langlands duality \cite{MTV2} --- it is not clear whether 3d spectral duality discussed here actually has anything to do with the spectral duality in {\it loc. cit.}
There are some hints suggesting that such a relation might exist. For instance, the Langlands duality that plays a key role in \cite{MTV2} in physics framework corresponds to the S-duality of the 4d gauge theory \cite{KW}. Similarly, we find that the relation between the effective twisted superpotential of 3d theories $\CT_{\text{here}}$ and $\CT_{\text{NS}}$ involves the $S$-transformation of 3d $\CN=2$ theories 
\begin{equation}
\W_\text{NS}(z;\zeta) \; = \; \W_\text{here}(z)+\log \zeta \log z \,.
\label{W_NS}
\end{equation}
To be a little more precise, the Nekrasov-Shatashvili duality relates a 3d $\CN=2$ theory $\CT_{\text{NS}}$ with a fixed rank of the gauge group to a particular ``sector'' of the corresponding integrable model.
For instance, in the case of the XXZ spin chain (to be discussed below) fixing the rank $N$ of the gauge group of the theory $\CT_{\text{NS}}$ corresponds to fixing the total number of excitations --- ``magnons'' --- in the XXZ spin chain. The spectral curve $\CV$, on the other hand, as well as its quantum version \eqref{quantumcurve} do not depend on this choice of sector.

\begin{figure}
\centering
\includegraphics{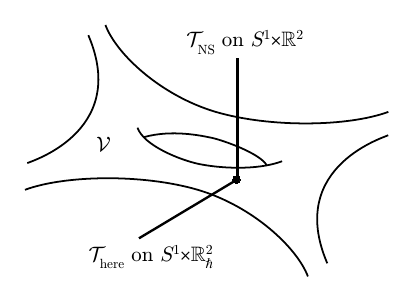}
\caption{Two rays depicting 3d theories $\mathcal{T}_\text{here}$ and $\mathcal{T}_\text{NS}$ supported on two different subspaces of the five-dimensional space-time $S^1 \times \R^2_{\hbar} \times \mathbb{R}^2$, where 5d theory with the Seiberg-Witten curve $\mathcal{V}$ lives. This brane interpretation makes it clear that $\CV$ plays the role of the SUSY parameter space for both 3d theories.}
\label{figure_curve-surf-op}
\end{figure}

Therefore, we conclude that the 3d $\CN=2$ theory $\CT_{\text{here}}$ defined by the dictionary \eqref{thedictionary} can only be dual to a {\it particular} version of the theory $\CT_{\text{NS}}$, {\it viz.} the most elementary version ({\it e.g.} with $N=1$, or a single magnon) that we call $\CT_{\text{NS}}^{\text{magnon}}$. Based on this observation and \eqref{W_NS} we, therefore, propose that 3d spectral duality relates 3d $\CN=2$ theory $\CT_{\text{here}}$
to the basic version of the Nekrasov-Shatashvili theory, $\CT_{\text{NS}}^{\text{magnon}}$:
\beq
\boxed{\phantom{\int} 3d~\text{theory}~\CT_{\text{NS}}^{\text{magnon}} \phantom{\int}}
\quad {S\text{-transformation} \atop \overleftrightarrow{\phantom{S-transformation}}} \quad
\boxed{\phantom{\int} 3d~\text{theory}~\CT_{\text{here}} \phantom{\int}}
\label{3dSduality}
\eeq
In what follows we justify this proposal and provide evidence to \eqref{W_NS} by considering concrete examples. Meanwhile, we point out that, in the brane construction of section~\ref{sect_branes}, the 3d spectral duality \eqref{3dSduality} corresponds to a rotation of the D3-brane / M2-brane by a $90$-degree angle,\footnote{We thank N.~Nekrasov for useful discussions on this point.} see Figure~\ref{figure_curve-surf-op}.

The simple relation by the $S$-transformation \eqref{W_NS} between $\W_\text{here}$ and $\W_\text{NS}$ in the abelian case (and in the limit $q\to 1$) is not unexpected since the brane model for $\CT_\text{NS}$ considered in \cite{DHL} is essentially the same (up to a rotation of the NS5$'$-brane by $\frac{\pi}{2}$) as the brane model for $\CT_\text{here}$ described in section~\ref{sect_branes_usual}. Further evidence for the interpretation of $\CT_{\text{here}}$ and $\CT_{\text{NS}}$ as effective world-volume theories on ``orthogonal'' surface operators in 5d has already been mentioned in the footnote~\ref{foot:adj}: in the framework of \cite{NS} the quantization parameter $\hbar = \log q$ is identified with the equivariant parameter of the $\Omega$-background in the plane {\it orthogonal} to the 3d space-time of the theory $\CT_{\text{NS}}$, whereas in the present framework $\hbar = \log q$ is the equivariant parameter of the $\Omega$-background {\it along} the 3d space-time of the theory~$\CT_{\text{here}}$.

For surface operators associated to knots \cite{DGH}, this agrees with the interpretation of the quantization parameter $q$ that turns classical (super) $A$-polynomial curves into quantum operators that annihilate polynomial and homological knot invariants, see {\it e.g.} \cite[sec.4]{FGS} for a detailed discussion and comparison with the Nekrasov-Shatashvili limit. Three-dimensional $\CN=2$ theories on such surface operators associated to knots will be the subject of section~\ref{sec:knots}.

Since turning on the $\Omega$-background of the form $(\hbar,0)$ corresonds to quantum deformation of the classical curve to the Baxter equation with parameter $q = e^\hbar$, one expects that introducing the general $\Omega$-background $(\epsilon_1,\epsilon_2)$ will trigger one-parameter refinement of the Baxter equation. It would be interesting to investigate this direction further.

\subsection{XXZ magnet and sinh-Gordon model}

In order to keep our discussion concrete and explicit, let us illustrate how 3d spectral duality \eqref{3dSduality} works in a large class of examples with Baxter equations of the form:
\begin{equation}
A(x) Q(qx) + D(x) Q(xq^{-1}) \; = \; T(x)Q(x)
\label{Baxtereq}
\end{equation}
where $A(x)$, $D(x)$, and $T(x)$ are some polynomials in $x$. In these models, $\log x$ is the ``true'' spectral parameter. Clearly, the Baxter equation \eqref{Baxtereq} can be written in the operator form \eqref{quantumcurve}--\eqref{xycomm} with
\beq
\hat{A}(\hat x,\hat y;q) \; = \; A(x) \hat y - T(x) + D(x) \hat y^{-1}
\eeq
and in the limit $q \to 1$ leads to the classical spectral curve $\CV$ defined by the equation:
\begin{equation}
A(x)y-T(x)+D(x)y^{-1} \; = \; 0 \,.
\label{BaxterClassical}
\end{equation}

This class of examples includes $sl(2)$ XXZ spin chain\footnote{Generalization to $sl(n)$ spin chains involves Baxter equations of $y$-degree $n$.} \cite{GGM2,MM5d,PronkoS,NS}, its modular version considered in \cite{TesBytsko}, the lattice version of the sinh-Gordon model, and other trigonometric / hyperbolic integrable systems. Thus, for the $sl(2)$ Heisenberg XXZ model of length $L$ both $A(x)$ and $D(x)$ are polynomials of degree~$L$:
\begin{equation}
A(x) \; = \; \prod_{j=1}^L(xe^{m_j/2}-e^{-m_j/2}) \,,
\label{AXXZ}
\end{equation}
\begin{equation}
D(x) \; = \; \prod_{j=1}^L(xe^{-\tilde{m}_j/2}-e^{\tilde{m}_j/2})
\label{DXXZ}
\end{equation}
and $T(x)$ is an eigenvalue of the transfer matrix $\hat T(x)$. Further specialization to
\beq
m_j = \tilde{m}_j = - 2 \pi b \sigma
\qquad j = 1, \ldots, L
\label{mmmodular}
\eeq
gives a modular XXZ magnet studied in \cite{TesBytsko}, and a similar specialization to
\begin{equation}
D(x)=A(x^{-1}) \; = \; \left( q^{\frac{1}{4}} x^{\frac{1}{2}} + e^{- 2 \pi b s} q^{- \frac{1}{4}} x^{-\frac{1}{2}} \right)^L
\end{equation}
gives the lattice sinh-Gordon model with $L$ sites. Here,\footnote{The definition of $q$ here differs from the definition in \cite{TesBytsko} by $q_\text{here}=q_\text{TB}^2$.} $q = e^{2i \pi b^2}$ and the parameter $\sigma := s + \frac{iQ}{2}$ (where $Q = b + b^{-1}$) determines the $q$-Casimir of the spin-$s$ principal series representation of $U_q (sl_2)$. This XXZ magnet is called modular \cite{TesBytsko} since it enjoys a remarkable $b \to b^{-1}$ duality.\\

Now, let us consider 3d $\CN=2$ theories $\CT_{\text{NS}}$ and $\CT_{\text{here}}$ associated to these integrable systems. Thus, the $N$-magnon sector of the $sl(2)$ XXZ spin chain of length $L$ is dual --- via the Nekrasov-Shatashvili duality \cite{NS} --- to the $U(N)$ gauge theory with one adjoint matter multiplet $\Phi$, $L$ chiral multiplets $Q_i$ in the fundamental representation of the gauge group, and $L$ chiral multiplets $\tilde Q_i$ in the anti-fundamental representation, see Table~\ref{tab:chargesNS}. We denote the mass parameters of these matter multiplets by $m_i$, $\tilde{m}_i$, and $m_\text{adj}$, respectively.
\begin{table}[htb]
\beq
\begin{array}{l@{\;}|@{\;}ccc}
& Q_{i = 1, \ldots, L} & \tilde Q_{i=1, \ldots, L} & \Phi \\\hline
U(N)_{\text{gauge}} & \Box & \overline{\Box} & \text{adj}  \\
U(1)_q            & -\frac{1}{2} & -\frac{1}{2} & +1
\end{array}
\qquad
\notag \eeq
\caption{Spectrum of the 3d $\CN=2$ theory $\CT_{\text{NS}}$ for the XXZ spin chain of length $L$.}
\label{tab:chargesNS}
\end{table}
\noindent
Upon compactification on a circle, one finds effective 2d theory with the twisted superpotential $\W_\text{NS}$ that depends on these mass parameters and the v.e.v.'s $\sigma_a$ of the scalar components in the vector multiplet. As explained in \cite{NS}, extremization of $\W_\text{NS}$ with respect to $\sigma_a$ leads to Bethe equations of the corresponding integrable system. For instance, starting with the 3d $\CN=2$ theory in Table~\ref{tab:chargesNS} one finds Bethe equations of the $sl(2)$ Heisenberg XXZ model of length $L$ with $N$ magnons.

In our present discussion, in particular in the relation \eqref{3dSduality}, we are mostly interested in the spectral curve~\eqref{Acurve} and its quantization, {\it i.e.} the Baxter equation~\eqref{quantumcurve}, which for the XXZ model in hand have the form~\eqref{Baxtereq} and~\eqref{BaxterClassical}, respectively. The way the Baxter equation arises from the theory $\CT_{\text{NS}}$ is a little indirect and goes via Bethe equations. Indeed, one can show that the latter are equivalent to the statement that a polynomial $Q (x) = \prod_{a=1}^N (x - e^{\sigma_a})$ obeys a $q$-difference equation~\eqref{Baxtereq} with $A(x)$ and $D(x)$ given by \eqref{AXXZ}-\eqref{DXXZ} and
\beq
\hbar = \log q = m_{\text{adj}}
\eeq
The theory $\CT_{\text{NS}}$ was interpreted in \cite{DHL,CDH} as a theory on a surface operator supported on $S^1 \times \mathbb{R}^2$ in 5d $\CN=2$ gauge theory on $S^1 \times \R^2_{\hbar} \times \mathbb{R}^2$, where $\hbar= \log q$ is the parameter of the Nekrasov-Shatashvili background \cite{NS4d}. In $\CT_\text{NS}$ it plays the role of the twisted mass for the axial $U(1)_q$ symmetry under which $\Phi$ has charge $+1$ and all (anti-)fundamental multiplets have charge $-\frac{1}{2}$, see Table~\ref{tab:chargesNS}. This $U(1)_q$ symmetry is a rotation symmetry in the two dimensional space {\it orthogonal} to the world-volume $S^1\times \mathbb{R}^2$ of the 3d $\CN=2$ gauge theory $\CT_{\text{NS}}$.

On the other hand, for the 3d $\CN=2$ theory $\CT_{\text{here}}$ defined by the rules \eqref{thedictionary} the quantum parameter $q$ is the equivariant parameter associated with the rotation symmetry {\it along} the two dimensions of the three-dimensional space-time (where the theory $\CT_{\text{here}}$ lives). Therefore, we interpret $\CT_{\text{here}}$ as a theory on the ``orthogonal'' surface operator supported on $S^1 \times_q \R^2$ in the same 5d gauge theory. Indeed, via M-theory lift \cite{Mtheory4d,Mtheory5d} 5d gauge theory on a circle is represented by M5-brane wrapped on the curve $A(x,y)=0$. In this setup, described in section \ref{sec:Mtheorybranes}, the surface operator is represented by a M2-brane \cite{ramified,AGGTV,DGOT} ending at a point on the curve $A(x,y)=0$.

To summarize, we list the similarities and differences between the 3d $\CN=2$ theories $\CT_{\text{here}}$ and $\CT_{\text{NS}}$ associated to the same integrable system:
\begin{itemize}
\item relation to integrable systems requires compactifying both theories $\CT_{\text{NS}}$ and $\CT_{\text{here}}$ on a circle;
\item extremization of the effective twisted superpotential most directly leads to Bethe equations in the case of $\CT_{\text{NS}}$,
and to the spectral curve / Baxter equation in the case of $\CT_{\text{here}}$;
\item closely related to the previous point, the theory $\CT_{\text{NS}}$ depends on $N$, whereas $\CT_{\text{here}}$ does not;
\item the quantum parameter $\hbar = \log q$ is related to the equivariant parameter along the 3d space-time of the theory $\CT_{\text{here}}$,
while in $\CT_{\text{NS}}$ it is identified with the mass of the adjoint multiplet $\Phi$.
\end{itemize}

In order to illustrate further the difference between the theories $\CT_{\text{here}}$ and $\CT_{\text{NS}}$, let us now discuss the theory $\CT_{\text{here}}$ that, via the rules \eqref{thedictionary}, corresponds to the modular XXZ magnet. In other words, $\CT_{\text{here}}$ is a 3d $\CN=2$ theory for which line operators obey a Ward identity \eqref{quantumcurve} identical to the Baxter equation \eqref{Baxtereq} with \eqref{AXXZ}-\eqref{DXXZ} and \eqref{mmmodular}. For simplicity, let us consider the basic case of $L=1$. In this case, the theory $\CT_{\text{here}}$ can be described as $\CN=2$ SQED with 2 chiral multiplets of charge $+1$ and masses $-\log\xi\pm\log\tilde{t}_{1}$ and 2 chiral multiplets of charge $-1$ and masses $- \log\xi\pm\log\tilde{t}_{2}$, see Table~\ref{tab:chargeshere}.
\begin{table}[htb]
\beq
\begin{array}{l@{\;}|@{\;}cccc@{\;}|@{\;}c}
& \phi_1 & \phi_2 & \phi_3 & \phi_4 & \text{~parameter} \\\hline
U(1)_{\text{gauge}} & 1 & 1 & -1 & -1 & z \\
U(1)_{\text{axial}} & 1 & 1 & 1 & 1 & \xi \\
U(1)_{t_1} & 1 & -1 & 0 & 0 & \tilde t_1 \\
U(1)_{t_2} & 0 & 0 & 1 & -1 & \tilde t_2
\end{array}
\qquad
\notag \eeq
\caption{Spectrum of the 3d $\CN=2$ theory $\CT_{\text{here}}$ for the modular XXZ magnet of length $L=1$.}
\label{tab:chargeshere}
\end{table}
\noindent
It is clear that the spectrum of this theory is quite different from that of theory $\CT_{\text{NS}}$ summarized in Table~\ref{tab:chargesNS}. The effective twisted superpotential $\W_\text{here}$ for the theory in Table~\ref{tab:chargeshere} has the form
\begin{equation}
\W_\text{here}(z,x)
\; = \; \dilog(z\xi^{-1}\tilde{t}_2)+\dilog(z\xi^{-1}\tilde{t}_2^{-1})
-\dilog(z\xi\tilde{t}_1)-\dilog(z\xi\tilde{t}_1^{-1})+\log x\log z-2\log z\log\xi
\end{equation}
where $\log x$ plays the role of the FI parameter. Extremizing with respect to $z$, which corresponds to the dynamical gauge symmetry of the theory $\CT_{\text{here}}$, one can check that the space of SUSY parameters $\CV$ is precisely the spectral curve~\eqref{BaxterClassical} of the modular XXZ magnet with $\pi b \sigma = - \log \xi$ and
\beq
T(x) \; = \; t_1 x - t_2 \,, \qquad \tilde{t_i}= \frac{t_i+\sqrt{t_i^2-4}}{2} \,.
\eeq
On the quantum level, the partition function of the theory on $S^1\times_q D$ is given by 
\begin{equation}
 Z^{\CT_\text{here}}_{S^1\times_q D}=\int \frac{dz}{z}\,
 \frac{\theta(z;q)\theta^2(x\xi;q)}{\theta(xz;q)\theta(x;q)\theta^2(\xi;q)}\,
 \frac{(qx\xi^{-1}\tilde{t_2};q)_\infty(qx\xi^{-1}\tilde{t_2}^{-1};q)_\infty}{(x\xi\tilde{t_1};q)_\infty (x\xi\tilde{t_1}^{-1};q)_\infty}
\end{equation}
and satisfies the Baxter equation (\ref{Baxtereq})\footnote{When the partition function on $S^1\times_q D$ has the form
\begin{equation}
Z(x)=\int \frac{dz}{z}\,
 \frac{\theta(z)\theta(x)}{\theta(xz)}\,
 Z^{S^{-1}}(z),
\end{equation}
i.e. is obtained by the $S$-transformation (\ref{S-transf-Z}) from $Z^{S^{-1}}$, the Ward identity 
\begin{equation}
\hat{A}(\hat{x},\hat{y})Z(x):= \sum_{m,n} A_{mn}\hat{x}^m\hat{y}^nZ(x)=0
\end{equation}
 can be rewritten as a Ward identity on $Z^{S^{-1}}$:
\begin{equation}
  \sum_{m,n} A_{mn}\hat{y}^{-m}\hat{x}^nZ^{S^{-1}}(x)=0.
\end{equation}
}. The field content of $\CT_\text{here}$ can be easily read off from this expression \cite{BDP}.

Let us note that the theory $\CT_\text{here}$ can in principle be constructed from the curve (\ref{BaxterClassical}) for any given polynomial $T(x)$. {}From the viewpoint of $\CT_\text{NS}$ the polynomial $T(x)$ is {\it a priori} unknown and should be determined. In particular it depends on the choice the gauge group of $\CT_\text{NS}$. In the simplest case $N=0$, which corresponds to the vacuum since $N$ is the number of magnons, we have $Q(x)=\text{const}$ and $T(x)$ can be determined from the following condition~\cite{DHL}:
\begin{equation}
A(x)+D(x)-T(x) \; = \; 0 \,.
\label{cond_on_T}
\end{equation}
The same condition can be obtained in the limit $q\rightarrow 1$ for general rank of the gauge group --- and, in particular, for $U(1)$ --- of $\mathcal{T}_\text{NS}$ assuming it is kept finite, {\it i.e.} $Q(x)$ is a polynomial of finite degree. From the condition (\ref{cond_on_T}) it follows that the spectral curve (\ref{spectral_curve}) is degenerate\footnote{In order to obtain a non-degenerate curve one has to consider 't Hooft-like limit for the rank of the gauge group~\cite{DHL}}:
\begin{equation}
A(x,y)= (1-1/y)(A(x)y-D(x))=0.
\label{spectral_curve}
\end{equation}
 The theory $\CT_\text{here}$ associated to the reduced curve
\begin{equation}
y=\frac{D(x)}{A(x)}=\prod_{j=1}^L\frac{xe^{-\tilde{m}_j/2}-e^{\tilde{m}_j/2}}{xe^{m_j/2}-e^{-m_j/2}}
\label{spectral_curve}
\end{equation}
contains $L$ chiral multiplets with charges $+1$ with respect to the global symmetry $U(1)_x$ and  $L$ chiral multiplets with charges $-1$. The corresponding effective twisted superpotential looks like
\begin{equation}
\W_\text{here}(x) \; = \; \sum_{j=1}^L\left[\dilog(x^{-1}e^{\tilde{m}_j})-\dilog(x^{-1}e^{-{m}_j})\right]
-\frac{1}{2}\log x\sum_{j}(m_j+\tilde{m}_j) \,.
\label{W_here}
\end{equation}
The effective twisted superpotential of $\CT_\text{NS}$ with the gauge group $U(1)_z$ and non-zero FI parameter $\log\zeta$ is related to it by the $S$-transformation~\eqref{W_NS}, up to a $z$-independent constant. There is no non-trivial contribution from the adjoint chiral multiplet in the abelian case.

Then, extremization of $\W_\text{NS}(z;\zeta)$ with respect to $z$ leads to Bethe equations in the case of a single Bethe root:
\begin{equation}
\zeta\prod_{j=1}^L\frac{ze^{-\tilde{m}_j/2}-e^{\tilde{m}_j/2}}{ze^{m_j/2}-e^{-m_j/2}} \; = \; 1 \,.
\label{single_Bethe_eq}
\end{equation}

\subsection{Trigonometric Ruijsenaars-Schneider system}

The spectral curve of the trigonometric Ruijsenaars-Schneider system \cite{Ruij,RS} (see also \cite{Nikita,BradenMMM} for appearance in the context of perturbative 5d Seiberg-Witten theory with adjoint matter) has the following form:
\begin{equation}
 y=\prod_{i=1}^L\frac{1-xe^{-m_i}}{1-xe^{-m_i}t}.
\label{RS-curve}
\end{equation}
The corresponding 3d theory $\CT_\text{here}$ is of type (\ref{W-ndSQED}) and contains $L$ chiral multiplets of charge $+1$ w.r.t. $U(1)_x$ and mass parameters $m_i$ and $L$ chiral multiplets of charge $-1$ and mass parameters $m_i-\log t$.  On the quantum level (\ref{RS-curve}) corresponds to the following difference equation:
\begin{equation}
 Q(qx)=\prod_{i=1}^L\frac{1-xe^{-m_i}}{1-xe^{-m_i}t}Q(x).
\label{RS-Baxtereq}
\end{equation}
It has the following simple solution:
\begin{equation}
 Q(x)=\prod_{i=1}^L\frac{(xe^{-m_i}t;q)_\infty}{(xe^{-m_i};q)_\infty},
 \label{RS-solution}
\end{equation}
given by an eigenvalue of the Baxter operator $\check{Q}(x)$ acting on space of symmetric polynomials in variables $\{e^{-m_i}\}$ constructed in \cite{GLO}. The eigenfunctions are given by Macdonald polynomials $P_\lambda(e^{-m_i};q,t)$ \cite{Macd}. From the point of view of $\CT_\text{here}$, (\ref{RS-solution}) is the partition function on $S^1 \times_q D$.


\section{Periods of (super) $A$-polynomial curves}
\label{sec:knots}

In our study of domain walls and line operators, the central role is played by the complex Lagrangian submanifold $\CV$ that we associate to a 3d $\CN=2$ theory $\CT$. As we explained earlier, $\CV$ has a simple interpretation as the space of SUSY parameters (FI terms and twisted masses) of the theory $\CT$ on a finite radius circle, and its ambient space \eqref{4d_moduli} can be identified with the moduli space of 4d $\CN=4$ gauge theory on the half-space $S^1_t \times \mathbb{R}^2 \times \mathbb{R}_+$ coupled to the 3d theory $\CT$ on the boundary.

Away from the boundary, this four-dimensional gauge theory is simply a 4d $\CN=4$ gauge theory compactified on a circle. Its physics is described by the effective 3d sigma-model, whose target space is the moduli space of (stable) Higgs bundles \cite{HMS,BJSV,SW3d}, the so-called Hitchin moduli space,
\begin{equation}
\CM_{4d} \; = \; \CM_H (G,C) \; \cong \; \CM_{\text{flat}} (G_{\C}, C) \,,
\label{MMMisoms}
\end{equation}
where in the last relation we used a well know fact that, in one of its complex structures, the Hitchin moduli space can be identified with the moduli space of flat connections on the Riemann surface $C$ for the complexified gauge group $G_{\C}$. Recently, this interpretation of the moduli space $\CM_{4d}$ played an important role in the gauge theory approach to the geometric Langlands program \cite{KW,ramified}, knot homologies \cite{GukovRTN,fiveknots}, and wall crossing phenomena \cite{GMN1}.

In our applications, using \eqref{MMMisoms} we can write
\begin{equation}
\CV \; \subset \; \CM_H (G,C) \; \cong \; \CM_{\text{flat}} (G_{\C}, C) \,,
\label{VinMM}
\end{equation}
where $C$ is a genus-1 Riemann surface and $G$ is a group of rank $n$ (typically, abelian). Moreover, for simplicity and concreteness we often focus on the case $n=1$, relevant to $G = U(1)$ or $G=SU(2)$. (The latter choice requires to include in \eqref{VinCC} a quotient by $\mathbb{Z}_2$, the Weyl group of $G = SU(2)$ or $G_{\C} = SL(2,\C)$.) Then, for suitable 3d $\CN=2$ theories, \eqref{VinMM} can be identified with the moduli space of classical solutions in $SL(2,\C)$ Chern-Simons theory on a 3-manifold $M$ with a toral boundary $C = \partial M = T^2$, which in general is also a complex Lagrangian submanifold of $\CM_{\text{flat}} (G_{\C}, C)$. In particular, when $M = S^3 \setminus K$ is a complement of a knot $K$ in a 3-sphere, the variety $\CV$ is the zero locus of the so-called $A$-polynomial \cite{CCGL} that determines the full quantum partition function of $SL(2,\C)$ Chern-Simons theory with a Wilson line supported on $K$ \cite{Apol,DGLZ}.

Based on this interpretation of $\CV$, it was conjectured \cite{DGH} that every knot $K$ (or, more generally, every 3-manifold with boundary) has a dual 3d $\CN=2$ theory $\CT_K$,
\begin{equation}
\text{knot}~K  \quad \leadsto \quad \text{3d}~\CN=2~\text{theory}~\CT_K \quad \leadsto \quad \text{SUSY parameter space}~\CV
\label{TKfromK}
\end{equation}
such that its SUSY parameter space $\CV$ can be identified with the moduli space of flat $G_{\C}$ connections on $M = S^3 \setminus K$, twisted superpotential $\tilde \CW$ with the classical Chern-Simons action \cite{DGSdual}, the supersymmetric partition functions \eqref{ZZZasymp} with variants of the complex Chern-Simons partition functions on $M$, {\it etc.} See \cite{DGG1} for a more complete dictionary. The 3d $\CN=2$ theory $\CT_K$ assigned to a knot $K$ can also be thought of as the effective theory obtained by reduction of the six-dimensional $(2,0)$ theory of $A_n$ type on $M$.

In the terminology of \cite{KW,ramified}, one can say that \eqref{VinMM} defines an $(A,B,A)$ brane supported on a submanifold $\CV$ in the Hitchin moduli space $\CM_H (G,C)$. Recently, it was realized that this $(A,B,A)$ brane and the setup \eqref{VinMM} admit a 2-parameter family of deformations (parametrized by complex variables $a$ and $t$) related to categorification of knot invariants \cite{FGP,FGSS},
$$
\text{knot}~K  \quad \leadsto \quad (a,t)\text{-deformation of}~\CV \subset \CM_H (G,C) \quad \leadsto \quad \text{colored HOMFLY homology}
$$
where the $(a,t)$-deformation of the space $\CV$ also can be interpreted as the moduli space of SUSY parameters in a 3d $\CN=2$ theory $\CT_K$ associated to a knot $K$ in a canonical way.\footnote{See also \cite{FGS,AV} for earlier developments and \cite{GSaberi} for a pedagogical introduction.}\\

In this section, our goal is to apply the results and lessons we learnt earlier to 3d $\CN=2$ theories $\CT_K$ and moduli spaces $\CV$ associated to knots. With some abuse of notations, we denote these by $\CT_K$ and $\CV$ even when we consider them in a more general, ``homological'' context that incorporates the $(a,t)$-deformation of the moduli spaces \eqref{VinMM}. In particular, for such a theory the corresponding curve \eqref{Acurve} is given by the zero locus of the super-$A$-polynomial,
\beq
\CV~: \quad A^{\text{super}} (x,y;a,t) \; = \; 0 \,,
\label{Asupercurve}
\eeq
whose explicit form is known for many knots \cite{FGP,Nawata,FGSS}.

Mathematically, the problem of computing periods of the 1-form $\log y \frac{dx}{x}$ on the super-$A$-polynomial curve \eqref{Asupercurve} is equivalent to a similar computation of masses of BPS states in 5d $\CN=2$ gauge theory on $\R^4 \times S^1$. On the one hand, such theories can be geometrically engineered via M-theory compactifications on toric Calabi-Yau 3-folds \cite{engineering,NLawrence}, which allows to identify their Seiberg-Witten curves with the geometry of mirror Calabi-Yau manifolds:
\begin{equation}
A(x,y;\{t_i\}) \; = \; 0 \,,
\label{Ati}
\end{equation}
where $A$ is polynomial in $x,y$ and in the complex structure parameters $t_i$. For a curve $\Cs$ of this form, the mirror map (between the complex moduli of $\Cs$ and the K\"ahler parameters of the Calabi-Yau 3-fold associated to the toric diagram $\Cs_\trop$) is trivial. Therefore, one can borrow many results and powerful methods on period computations, including the Picard-Fuchs equations {\it etc.}, and in the end specialize to particular values of moduli $t_i$ 
that make the 5d Seiberg-Witten curve identical to the super-$A$-polynomial curve \eqref{Asupercurve}. Put differently, super-$A$-polynomial curves \eqref{Asupercurve} can be viewed as special cases of 5d Seiberg-Witten curves (or mirror curves in the geometric engineering of such theories) with parameters $\{t_i\}=\{a,t\}$.

In a different direction, many 5d $\CN=2$ gauge theories correspond to relativistic or trigonometric integrable systems \cite{GKMMM,Nikita,GGM2,MM5d,BradenMMM}. This duality identifies the Seiberg-Witten curves of 5d theories with spectral curves of the corresponding integrable models and is consistent with our earlier discussion, where the same curve $\Cs$ was also identified with the space of SUSY parameters on a surface operator in 5d theory.

Returning to 3d $\CN=2$ theories $\CT_K$ associated to knots as in \eqref{TKfromK}, the effective twisted superpotential is expected \cite{FGP} to be of the form (\ref{W-general-lagrangian}). As was previously noted, in such cases all periods have the form (\ref{periods_moduli}), namely for \textit{any} cycle $\gamma\in H_1(\Cs)$:
\begin{equation}
 \Delta_{\gamma}\W (x)=\int_{\gamma}\log y\, d\log x
 =2\pi^2 c_{\gamma}+2\pi i\sum_{i}q_{\gamma,i}\log t_i+2\pi i n_{\gamma} \log x,
 \qquad q_{\gamma,i},n_\gamma,c_\gamma\in\mathbb{Z}
 \label{periods_lagrangian_0}
\end{equation}
For a given $\gamma$, one can make a choice of ``polarization,'' {\it i.e.}
\begin{equation}
 \left(\begin{array}{c} \log\tilde{y}_\gamma \\ \log\tilde{x}_\gamma\end{array}\right)=M_\gamma
  \left(\begin{array}{c} \log{y} \\ \log{x}\end{array}\right),\qquad M_\gamma\in SL(2,\mathbb{Z})
\end{equation}
such that the period
\begin{equation}
 \int_{\gamma}\log \tilde{y}_\gamma\, d\log \tilde{x}_\gamma
 =2\pi^2 \tilde{c}_{\gamma}+2\pi i\sum_{i}q_{\gamma,i}\log t_i,
 \qquad q_{\gamma,i},c_\gamma\in\mathbb{Z}
 \label{periods_lagrangian}
\end{equation}
does not depend on the position of the beginning / end-point of the contour.

The fact that there are no non-trivial periods of the form (\ref{periods_faces}) is related to the condition that the curve is {\it quantizable} \cite{abmodel}. From (\ref{periods_moduli}) one can deduce that for the curve to be quantizable it is necessary that
\begin{equation}
t_i=\pm q^{\ell_i},\qquad \ell_i\in \mathbb{Z} \label{params_q_cond}
\end{equation}
where $q=e^\hbar$. If the first and the third terms in (\ref{periods_faces}) are non-zero, then in general the quantization condition
is not satisfied for these periods. However, when (\ref{periods_faces_cond}) holds, the conditions (\ref{params_q_cond}) are also sufficient for quantizability of the curve $\CV$. Among other things, this explains the observations \cite{FGS,FGP} that super-$A$-polynomial curves \eqref{Asupercurve} seem to be quantizable when the refinement parameter $t$ is equal to $q^{\ell}$ for some integer $\ell \in \mathbb{Z}$. This also gives evidence to the conjecture \cite{AV} that $t=-1$ specializations of \eqref{Asupercurve} provide infinitely many mirrors of the resolved conifold with the K\"ahler parameter $\log a$. Indeed, eq. (\ref{periods_lagrangian}) tells us that all periods are proportional to $\log a$.


\subsection{Trefoil knot}

The simplest non-trivial example of the theory $\CT_K$ is a 3d $\CN=2$ theory \eqref{TKfromK} associated to the trefoil knot. It is a $U(1)$ gauge theory with the effective twisted superpotential \cite{FGP}:
\begin{multline}
\W(z;x)=-\pi^2/6+(\log z+\log a)\log x+2\log t\log z+\\
\dilog(x/z)-\dilog(x)+\dilog(-at)-\dilog(-atz)+\dilog(z)
\end{multline}
that leads to a genus-1 curve \eqref{Asupercurve} defined by the zero locus of the super-$A$-polynomial:
\beq
A^{\text{super}} (x,y; a,t) = c_0 + c_1 y + c_2 y^2 \,.   \label{trefoil-curve}
\eeq
In other words, the super-$A$-polynomial of the trefoil knot is a quadratic polynomial in $y$, whose coefficients are polynomial expressions in $x$, $a$, and $t$:
\bea
c_0 & = & a^2 t^4 (x-1) x^3 \nonumber \\
c_1 & = & -a\big( 1 - t^2 x + 2 t^2 (1 + a t) x^2 + a t^5 x^3 + a^2 t^6 x^4 \big) \nonumber \\
c_2 & = & 1 + a t^3 x   \nonumber
\eea
From \eqref{trefoil-curve} it is easy to see that the curve $A^{\text{super}} (x,y; a,t) = 0$ can be identified with the Seiberg-Witten curve of 5d $\CN=2$ gauge theory with gauge group $G=U(4)$ and $N_f=2$ fundamental matter multiplets \cite{Nikita}\footnote{There are many equivalent ways to write this curve, {\it e.g.} introducing $x = e^{\lambda}$ and making a simple change of variables it can be brought to a typical ``trigonometric form''
$$
y \sinh \frac{\lambda - m_1}{2} + \Lambda^{2}\frac{ \sinh \frac{\lambda - m_2}{2}}{y}
\; = \; \prod_{j=1}^4 \sinh \frac{\lambda - a_j}{2}
$$
}:
\bea
\alpha (x - e^{m_1}) y + \beta\,\frac{x^3 (x - e^{m_2})}{y}
\; = \; P(x) & = & \prod_{j=1}^4 (x - e^{a_j}) \\
& = & x^4 + u_1 x^3 + u_2 x^2 + u_3 x + u_4  \nonumber
\eea
provided we identify
\bea
e^{m_1} & = & -at^{-3} \nonumber \\
e^{m_2} & = & 1 \nonumber \\
u_1 & = & a^{-1}t^{-1} \nonumber \\
u_2 & = & 2 a^{-2}t^{-4} (1 + a t) \nonumber \\
u_3 & = & -a^{-2} t^{-4} \nonumber \\
u_4 & = & a^{-2} t^{-6}  \nonumber
\eea

\begin{figure}
\centering
\includegraphics{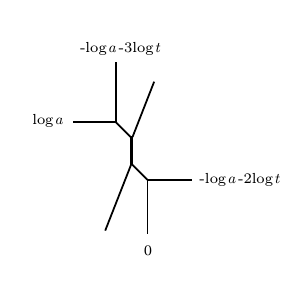}
\caption{The tropical limit of the curve for the trefoil knot. The thick vertical part in the middle consists of two edges of multiplicity two coinciding in the tropical limit.}
\label{figure_trefoil-trop}
\end{figure}

The tropical limit of the curve $A^{\text{super}} (x,y; a,t) = 0$ is shown in Fig. \ref{figure_trefoil-trop}. The graph has one face which becomes degenerate in the tropical limit ({\it i.e.} has zero area). This face is bounded by two vertical edges of multiplicity two. And the fact that it has vanishing area corresponds to the fact that there are no non-trivial periods of the form~(\ref{periods_faces}). Using the basic rules (\ref{mon_rules}) one can compute periods for a set of cycles that form a basis in $H_1(\Cs)$:
\begin{equation}
\begin{array}{rlcrl}
\mon_{\gamma_1}\W =& -2\pi i\log x-2\pi i\log a-6\pi i\log(-t),
&\qquad & c_*\gamma_1= & (-1,0), \\
\mon_{\gamma_2}\W =& 6\pi i\log x+2\pi i\log a+8\pi i\log(-t),
&\qquad & c_*\gamma_2= & (3,1),  \\
\mon_{\gamma_3}\W =& 2\pi i\log a,
&\qquad & c_*\gamma_3= & (0,1),  \\
\mon_{\gamma_4}\W =& 2\pi i\log x,
&\qquad & c_*\gamma_4= & (1,0), \\
\mon_{\gamma_5}\W =& -2\pi i\log a-4\pi i\log(-t),
&\qquad & c_*\gamma_5= & (0,1),  \\
\mon_{\gamma_6}\W =& 6\pi i\log x+4\pi i\log a +6\pi i\log(-t),
&\qquad & c_*\gamma_6= & (3,1), \\
\mon_{\gamma_7}\W =& -4\pi i\log x-2\pi i\log a-6\pi i\log(-t),
&\qquad & c_*\gamma_7= & (-2,0),  \\
\mon_{\gamma_8}\W =& -2\pi i\log a-2\pi i\log(-t),
&\qquad & c_*\gamma_8= & (0,0) \\
\end{array}
\end{equation}
To be more specific, the cycles $\gamma_1,\ldots,\gamma_6$ are associated to external legs, the cycle $\gamma_7$ is associated to one of the finite vertical edges and the cycle $\gamma_8$ is associated to the degenerate face.

\subsection{Figure-eight knot}

The non-deformed $A$-polynomial for the figure-eight knot (see {\it e.g.} \cite{CCGL,Apol}):
\begin{equation}
A(x,y) \; = \; (y^2+1)x^2-y(1-x-2x^2-x^3+x^4)
\end{equation}
leads to a curve \eqref{Acurve} of genus one. It is a specialization of \eqref{Asupercurve} at $a=1$ and $t=-1$, which already is rich enough and interesting enough. The corresponding theory $\CT_K$ has the following effective twisted superpotential \cite{FGP}:
\begin{equation}
\W \; = \; -\dilog(xz)+\dilog(x/z)-\log x\log z \,.
\end{equation}
There are a total of six independent cycles with the following periods and charges:
\begin{equation}
\begin{array}{rlcrl}
\mon_{\gamma_1}\W =& 4\pi i\log x-4\pi^2,
&\qquad & c_*\gamma_1= & (2,1), \\
\mon_{\gamma_2}\W =& -4\pi i\log x+4\pi^2 ,
&\qquad & c_*\gamma_2= & (-2,1),  \\
\mon_{\gamma_3}\W =& 4\pi i\log x-4\pi^2 ,
&\qquad & c_*\gamma_3= & (2,-1),  \\
\mon_{\gamma_4}\W =& -4\pi i\log x+4\pi^2 ,
&\qquad & c_*\gamma_4= & (-2,-1), \\
\mon_{\gamma_5}\W =& 2\pi i\log x,
&\qquad & c_*\gamma_5= & (1,0),  \\
\mon_{\gamma_6}\W =& -4\pi^2,
&\qquad & c_*\gamma_6= & (0,1), \\
\end{array}
\label{figure8-charges}
\end{equation}
where the cycles $\gamma_1,\ldots,\gamma_4$ are associated to the tentacles of the amoeba.

Following \cite{BorotE}, we note that after a rational change of variables the elliptic curve $A(x,y)=0$ in this example can be written in the standard Weierstrass form
\beq
y^2 = 4 x^3 - g_2 x - g_3
\label{Weierstrass}
\eeq
with $g_2 = \frac{x}{12}$ and $g_3 = - \frac{161}{216}$.
The same trick can be used to bring $A$-polynomial curves for many other knots, including $9_{35}$, $9_{48}$, and $10_{139}$ discussed in \cite{BorotE}, into the Weierstrass form. It is important to keep in mind, though, that the rational change of variables which does that in general transforms $\log y \frac{dx}{x}$ into some other (less canonical) 1-form. Therefore, even though \eqref{Weierstrass} is reminiscent to the form of the Seiberg-Witten curve in $SU(2)$ gauge theory, the computation of periods that arise from application to knots are rather different to computation of the periods of the Seiberg-Witten differential.


\subsection{``Incompressible operators''}

We conclude this section by extending the dictionary \cite{DGG1} of 3d/3d duality to include {\it incompressible surfaces}, which play an important role in low-dimensional topology \cite{ARindecomp}, but so far managed to escape attention of physicists.

In fact, the discussion of incompressible surfaces is unavoidable in the present context where the algebraic curve $A(x,y)=0$ plays a central role. Indeed, on the one hand, this curve contains a great deal of information about the spectrum of incompressible surfaces in a 3-manifold $M$ \cite{CCGL}. And, on the other hand, as we explained in section \ref{sect_defects}, the same algebraic curve --- interpreted via 3d/3d duality as a space of SUSY parameters --- encodes the spectrum of line operators in the corresponding 3d $\CN=2$ theory $\CT_M$. When combined together, these two relations imply that there should exist a direct link between incompressible surfaces in $M$ and line operators in the theory $\CT_M$ and, as we explain below, it does indeed.

By definition, a proper embedding of a connected orientable surface $\Sigma \hookrightarrow M$ into a 3-manifold $M$ with boundary\footnote{which in most of our applications we take to be a knot complement $M = S^3 \setminus K$} is called incompressible if the induced map
\beq
\pi_1 (\Sigma) \to \pi_1 (M)
\eeq
is injective. Another, closely related definition of an incompressible surface $\Sigma \ne S^2$ is as a surface that does not bound any compressing disk, {\it i.e.} any disk $D \subset M$ such that the loop $\partial D$ does not bound a disk in $\Sigma$.

\begin{figure}
\center{\includegraphics[width=4.5in]{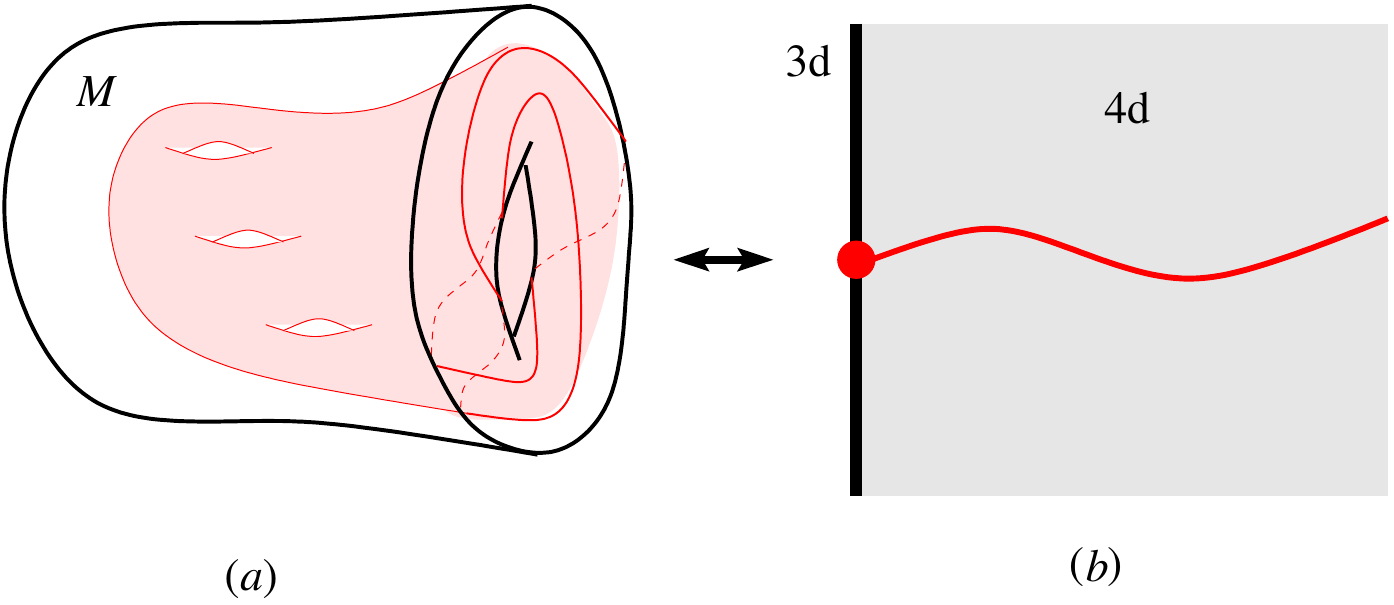}}
\caption{Incompressible surface $(a)$ in a 3-manifold $M$ via 3d/3d duality corresponds to $(b)$ a line operator in 4d gauge theory that ends at a local operator in the corresponding 3d boundary theory $\CT_M$.\label{fig:indecomp}}
\end{figure}

An important characteristic of incompressible surfaces is the {\it boundary slope}. It turns out to have a nice physical interpretation and is defined as follows. An incompressible surface $(\Sigma, \partial \Sigma)$ gives rise to a collection of parallel simple closed loops in $T^2 = \partial M$, see Figure \ref{fig:indecomp}$a$. Choose one such loop and write its homology class as\footnote{The origin of the extra factor of 2 in the following formulas is that the variable $x$ that we use in A-polynomials differs from the canonical one used in the Knot theory literature by $x_\text{here}=x_\text{canonical}^2$. }
\beq
(\text{longitude})^{2n} \; (\text{meridian})^m \quad \leadsto \quad \text{boundary slop} \; = \; \frac{2n}{m} \,.
\label{bslopedef}
\eeq
Then, the boundary slope of $(\Sigma, \partial \Sigma)$ is defined as a rational number $\frac{2n}{m}$.
For any compact orientable irreducible 3-manifold $M$ with a toral boundary the spectrum of boundary slopes consists of only finitely many values of $2n/m$ \cite{Hatcher}. Particular examples of values $(n,m)$ are given by the charges associated to the tentacles of the amœba associated to the A-polynomial.  For example, in Table~\ref{fig8incomp} we list all types of oriented incompressible surfaces for the figure-eight knot complement with their boundary slopes \cite{Thurston,HatcherT}.
\begin{table}[ht]
\centering
\renewcommand{\arraystretch}{1.3}
\begin{tabular}{|@{\quad}c@{\quad}|@{\quad}c@{\quad}|@{\quad}c@{\quad}| }
\hline  Name & Topology & Boundary Slope \\
\hline
\hline $\Sigma_1$ & $T^2 \setminus \text{disk} = $ Seifert surface & $0$ \\
\hline $\Sigma_2$ & $T^2 \setminus 2~\text{disks}$ & $\pm 4$ \\
\hline $\Sigma_2'$ & $T^2 \setminus 2~\text{disks}$ & $\pm 4$ \\
\hline
\end{tabular}
\caption{The spectrum of incompressible surfaces and boundary slopes for the figure-8 knot complement (cf. (\ref{figure8-charges})).}
\label{fig8incomp}
\end{table}

In order to explain the promised relation between incompressible surfaces and line operators, as usual in the discussion of 3d/3d duality \cite{DGGvirtue,DGH,HosomichiLP,TerashimaY,DGG1,CCV} let us consider the six-dimensional $(2,0)$ theory on $(S^1_t \times_q \R^2) \times M$. Since near the boundary $M$ looks like $T^2 \times \R_+$, after dimensional reduction on $M$ one finds a 4d $\CN=4$ gauge theory on half-space coupled to 3d $\CN=2$ theory $\CT_M$ on the boundary that we discussed in section \ref{sect_defects}.

Now, let us incorporate incompressible surfaces by introducing codimension-4 defects in 6d $(2,0)$ theory that can be interpreted as end-points of M2-branes ending on M5-branes and that give rise to surface operators \cite{ramified}. When such a surface operator wraps a non-trivial cycle of $T^2 = \partial M$ it gives rise to a line operator in the resulting 4d $\CN=4$ gauge theory with gauge group $U(1)$. Specifically, the electric charge $n$ of this line operator and its magnetic charge $m$ are determined by the homology class of the curve in $T^2$, see \cite{Henningson,DMO,AGGTV,DGOT} and discussing in section \ref{sect_defects}. Therefore, what we called $n$ and $m$ in the definition \eqref{bslopedef} of the boundary slope are precisely electric and magnetic charges of line operator in 4d abelian gauge theory.

This would be the end of the story if our 3-manifold had the form $M = \R \times T^2$, but since it is ``capped off'' as illustrated in Figure \ref{fig:indecomp}$a$, the resulting 4d gauge theory lives on a half-space with a 3d $\CN=2$ theory $\CT_M$ on the boundary, see Figure \ref{fig:indecomp}$b$. Correspondingly, via 3d/3d duality the codimension-4 surface operator supported on $\Sigma \subset M$ maps into a line operator that terminates at a local operator $\CO$ on the boundary. When $\Sigma$ is incompressible it is natural to refer to the corresponding operator $\CO$ in 3d $\CN=2$ theory $\CT_M$ as the ``incompressible operator.''

According to the above mentioned result \cite{Hatcher}, ``incompressible operators'' in 3d $\CN=2$ theory $\CT_M$ have a finite spectrum of values $\frac{2n}{m}$. For a signature of such operators one can look at the superconformal index $\CI (m,n)$ in the charge sector $(m,n)$ \cite{DGGindex}. For example, for the figure-eight knot the index $\CI_{\mathbf{4_1}} (m,n)$ contains the following data
\beq
\CI_{\mathbf{4_1}} (1,2) =
\CI_{\mathbf{4_1}} (1,-2) =
\CI_{\mathbf{4_1}} (-1,-2) =
\CI_{\mathbf{4_1}} (-1,2) = q^{5/2}+q^{7/2}-q^{11/2} + \dots \,,
\eeq
which should be compared with the values of the boundary slope $\pm 4$ of incompressible surfaces in Table~\ref{fig8incomp}.


\acknowledgments{We thank C.~Beem, A.~Bytsko, T.~Dimofte, A.~Gorsky, S.~Nawata, N.~Nekrasov, S.~Shatashvili, P.~Su{\l}kowski, R.~van der Veen for useful discussions on related topics.
The work of A.G. is supported in part by the John A. McCone fellowship and by  DOE Grant DE-FG02-92-ER40701. The work of S.G. is supported in part by DOE Grant DE-FG03-92-ER40701FG-02 and in part by NSF Grant PHY-0757647. The work of P.P. is supported in part by the Sherman Fairchild scholarship and by NSF Grant  PHY-1050729. Opinions and conclusions expressed here are those of the authors and do not necessarily reflect the views of funding agencies.}

\appendix

\section{\label{example-table}Basic examples}
The following table summarizes some basic examples of effective twisted superpotentials, corresponding curves, periods and associated charges. Cycles around singular points $(x_0,y_0)$ are marked by $x\approx x_0,y\approx y_0$.

{\footnotesize
\begin{center}
\begin{tabular}{|m{0.45\textwidth}|m{0.3\textwidth}|m{0.15\textwidth}|}
\hline
\vs
$\W$, curve & e/m charges

$(\Delta\log y,\Delta\log x)/2\pi i$ & $\Delta\W$ \\
\hline
\hline
\vs
$\W=0$

\vs
$y=1$
&
$(0,1)$
&
0 \\
\hline
\vs
$\W=k\log z\log x$

\vs
$x^k=1$
&
$k$ copies of $(1,0)$
&
$2\pi i\log x$ \\

\hline
\vs
$\W=\frac{k}{2}(\log x)^2$

\vs
$y=x^k$
&
$(k,1)$
&
$2\pi ik\log x-2\pi^2k$ \\
\hline
\vs
$\W=\frac{A}{2}(\log z)^2+B\log x\log z+\frac{C}{2}(\log x)^2$

\vs
$y^A=x^{AC-B^2}$
&
$\left( \begin{array}{ccc}
0 & 0 & 1 \\
0 & 1 & 0 \end{array} \right) \ker \left( \begin{array}{ccc}
A & B & 0 \\
B & C & -1 \end{array} \right)$
&
$(\Delta\log y)\log x$

$(\mod 2\pi^2)$  \\
\hline
\vs $\W=\beta\dilog(x)+\frac{k}{2}(\log x)^2 $

\vs $y=(1-x)^{-\beta}x^k$
&
$(k,1),\; x\approx 0,y\approx 0$

$(-\beta,0),\; x\approx 1,y\approx\infty$
&
$2\pi ik\log x-2\pi^2 k$

$-2\pi i \beta \log x$ \\
\hline
\vs $\W=\dilog(xz)+\log x\log z$

\vs $y=(1-x)/x^2$
&
$(-2,1),\; x\approx 0,y\approx \infty$

$(-1,1),\; x\approx \infty,y\approx 0$
&
$-4\pi i\log x+4\pi^2$

$-2\pi i \log x$ \\
\hline
\vs $\W=\dilog(x)+\dilog(z)+2\log x\log z$

\vs $y=(1-x)(1+x)^2$
&
$(1,1),\; x\approx 1,y\approx 0$

$(2,1),\; x\approx -1,y\approx 0$
&
$2\pi i\log x-16\pi^2$

$4\pi i \log x-16\pi^2$ \\
\hline
\vs figure-eight knot

\vs $\W=-\dilog(xz)+\dilog(x/z)-\log x\log z $

\vs $(y^2+1)x^2-y(1-x-2x^2-x^3+x^4)=0$
&
$(2,1),\; x\approx 0,y\approx 0$

$(-2,1),\; x\approx 0,y\approx \infty$

$(2,-1),\; x\approx\infty, y\approx 0$

$(-2,-1),\; x\approx\infty,y\approx\infty$

$(1,0)$

$(0,1)$
&
$4\pi i\log x-4\pi^2 $

$-4\pi i\log x+4\pi^2 $

$4\pi i\log x-4\pi^2 $

$-4\pi i\log x+4\pi^2 $

$2\pi i\log x$

$-4\pi^2$
 \\
\hline
\end{tabular}
\end{center}
}

\bibliography{./files/periodsref}
\bibliographystyle{./files/JHEP_ref}
\end{document}